\documentclass[12pt]{article}
\usepackage{amsmath, amssymb, amsfonts, ntheorem}
\usepackage{graphicx, placeins}
\usepackage{enumerate, enumitem}
\usepackage{natbib} 
\usepackage{xr-hyper} 
\usepackage{url, hyperref} 
\usepackage{algorithm, algorithmic}
\usepackage[font=small,skip=0pt]{caption}
\usepackage{booktabs, multirow}
\usepackage{float}
\newcommand*{\E}{\mathrm{E}}
\newcommand*{\Var}{\mathrm{Var}}
\newcommand*{\Cov}{\mathrm{Cov}}
\newcommand*{\COV}{\mathbf{Cov}}
\newcommand*{\SE}{\mathrm{SE}}
\newtheorem*{proof}{\textbf{Proof}}
\newtheorem{proposition}{Proposition}
\newtheorem{corollary}{Corollary}
\newtheorem{lemma}{Lemma}

\hypersetup{
  colorlinks   = true, 
  urlcolor     = black, 
  linkcolor    = black, 
  citecolor   =  black, 
  filecolor    = black 
}

\begin{document}

\def\spacingset#1{\renewcommand{\baselinestretch}%
{#1}\small\normalsize} \spacingset{1}


\title{\bf Active sampling: A machine-learning-assisted framework for finite population inference with optimal subsamples}

\newcommand\authors{Imberg, Yang, Flannagan and Bärgman}

\date{} 


\author{Henrik Imberg$^{1}$, Xiaomi Yang$^{2}$, Carol Flannagan$^{2,3}$, Jonas Bärgman$^{2}$ \medskip \\
\small
$^{1}$Department of Mathematical Sciences, \\
\small
Chalmers University of Technology and University of Gothenburg \medskip \\
\small
$^{2}$Division of Vehicle Safety, Chalmers University of Technology \medskip \\
\small
$^{3}$University of Michigan Transportation Research Institute}
\maketitle

\bigskip
\begin{abstract}
    Data subsampling has become widely recognized as a tool to overcome computational and economic bottlenecks in analyzing massive datasets. We contribute to the development of adaptive design for estimation of finite population characteristics, using active learning and adaptive importance sampling. We propose an active sampling strategy that iterates between estimation and data collection with optimal subsamples, guided by machine learning predictions on yet unseen data. The method is illustrated on virtual simulation-based safety assessment of advanced driver assistance systems. Substantial performance improvements are demonstrated compared to traditional sampling methods. 
\end{abstract}


\newpage
\spacingset{2} 
\section{Introduction}
\label{sec:intro}

We consider a deterministic computer simulation experiment which for a given input $\boldsymbol{z}$ returns a fixed output $\boldsymbol{y}$. The input space is assumed to be discrete and the simulation experiment hence characterized by the set of complete input-output pairs $\{(\boldsymbol{z}_i, \boldsymbol{y}_i)\}_{i = 1}^N$, where $N$ is the size of the experiment. The aim our experiment it to calculate a characteristic 
\begin{equation}
\label{eq:theta}
\theta = h(\boldsymbol{t}_{\boldsymbol{y}}), \quad \boldsymbol{t}_{\boldsymbol{y}} = \sum_{i=1}^N \boldsymbol{y}_i,
\end{equation}
for some differentiable function $h:\mathbb{R}^d \to \mathbb{R}$ and $d$-dimensional vector of totals $\boldsymbol{t}_{\boldsymbol{y}}$. Examples of such a characteristic include, e.g., a total, mean, ratio, or correlation coefficient. This is also known as a finite population inference problem \citep{Beaumont2022}. We further assume that $N$ is large, as is the computational cost of each single experiment, rendering complete enumeration unfeasible. In such circumstances, researches often resort to subsampling.

Subsampling methods have seen a huge increase in popularity over the past few years across many different areas of statistics. For instance, \citet{Ma2015, Ma2022} introduced leverage sampling for big data regression, which subsequently inspired similar developments for logistic regression \citep{Wang2018, Yao2019} generalized linear models \citep{Ai2021_regression, Yu2022}, and quantile regression \citep{Ai2021_quantile, Wang2021}. Similarly, \cite{Dai2022} developed an optimal subsampling method for regression using a minimum energy criterion. Sometimes subsampling is induced by economical rather than computational constraints. In this setting, \citet{Imberg2022} developed an optimal subsampling method for two-phase sampling experiments. A similar measurement-constrained experiment problem was addressed by \citet{Zhang2021} using a sequential subsampling procedure and by \citet{Meng2021} using a space-filling Latin hypercube sampling method.

For computer simulation experiments, subsampling methods using adaptive design for Gaussian process response surface modeling are commonly employed. Together with active learning and Bayesian optimization, this provides a powerful framework for computer experiment emulation \citep{Gramacy2015, Sun2017, Lei2021, Lim2021}.  Another popular approach is model-free space-filling methods using, e.g., Latin hypercube sampling designs \citep[see, e.g.,][]{Cioppa2007, Zhang2023, Zhou2023}. Others have utilized methods based on optimal transport, e.g., for kernel density estimation \citep{zhang2023optimal}. For estimating a simple statistic, such as a mean or ratio, however, importance sampling and adaptive importance sampling remains prominent \citep{bucher1988adaptive, Oh1992, Feng2021}. Importance sampling is widely known, easy to implement, and provides consistent estimates under minimal assumptions \citep{Fishman1996, Fuller2011}. Some recent developments include adaptive importance sampling for quantile estimation \citep{Pan2020} and online monitoring of data streams \citep{liu2015adaptive, xian2018nonparametric}.

There has also been a considerable interest in subsampling and adaptive design in machine learning, particularly in the context of active learning \citep{MacKay1992, Cohn1996, Settles2012}. Adaptive importance sampling methods for active learning were developed in, e.g., \citet{Bach2007}, \citet{Beygelzimer2009} and \citet{Imberg2020}. Active learning has also been utilized for deep learning \citep{ren2021survey}, Gaussian processes \citep{Sauer2023} and adaptive design of experiments \citep{lookman2019active, sun2021adaptive}, to mention a few.

Returning to the finite population inference problem \eqref{eq:theta}, this is a classical problem in statistics and hence has achieved considerable attention over the years, particularly in the survey sampling literature. Common approaches to estimation include importance sampling methods and/or using estimators that utilize information of known auxiliary variables to improve estimator efficiency \citep[see, e.g.,][]{Cassel1976, Deville1992, Kott2016, Ta2020}. Methods utilizing machine learning in survey sampling have just recently begun to emerge \citep{Breidt2017, Kern2019, McConville2019, Sande2021}. Although there has been a substantial amount of work on subsampling and adaptive design in the statistical literature, there is to our knowledge little done at the intersection of machine learning and adaptive design for the finite population inference problem \eqref{eq:theta}.

\paragraph{Contributions}
\label{sec:contributions}

To fill the gap in adaptive design and machine learning for finite population inference, we propose an active sampling strategy for estimation of finite population characteristics. Our method iterates between estimation and data collection with optimal subsamples, guided by machine learning predictions on yet unseen data. The proposed sampling strategy interpolates in a completely data-driven manner between simple random sampling when no auxiliary information is available and optimal importance sampling as more information is acquired. Consistency and asymptotic normality of the active sampling estimator is established using martingale central limit theory. Methods for variance estimation are proposed and conditions for consistent variance estimation presented.

\paragraph{Outline}
\label{sec:outline}

The structure of this paper is as follows: We start by presenting a motivating example in crash-causation-based scenario generation for virtual vehicle safety assessment in Section \ref{sec:intro_application}. Mathematical preliminaries and notation is introduced in Section \ref{sec:sampling}. In the end of this section we also derive an optimal importance sampling scheme for estimating a finite population characteristic while accounting for uncertainty in the study variables of interest. This is then incorporated in the active sampling algorithm proposed in Section \ref{sec:active_sampling_intro}. An empirical evaluation on simulated data is conducted in Section \ref{sec:experiments} and application to virtual vehicle safety assessment in Section \ref{sec:application}. Additional theoretical results and proofs are provided in Appendix \ref{appendix:theory}.

\section{Motivating example}
\label{sec:intro_application}

Traffic safety is a problem worldwide \citep{road_safety_2018}. Safety systems have been developed to improve traffic safety and  have shown the potential to avoid or mitigate crashes. However, when developing both advanced driver assistance systems and automated driving systems, there is a need to assess the impact on safety of the systems before they are on the market. One way to do that is by running virtual simulations comparing the outcome of simulations both with and without a specific system \citep{seyedi2021, leledakis2021method}.

We consider a virtual simulation experiment based on a glance-and-deceleration crash-causation model where a driver's off-road glance behavior and braking profile are represented by discrete (empirical) probability distributions, using a similar setup as in \citet{bargman2015does} and \citet{lee2018safe}. The outcome of the simulations is a distribution of impact speeds of all the crashes generated by all combinations of the eyes-off-road glace duration and the maximum deceleration during braking. Here “all combinations” is the problem. Complete enumeration becomes practically unfeasible in high-dimensional (many parameters varied) or high-resolution (many levels per parameter) settings, and subsampling is inevitable.

\begin{figure}[htb!]
    \centering
    \includegraphics[scale = 1]{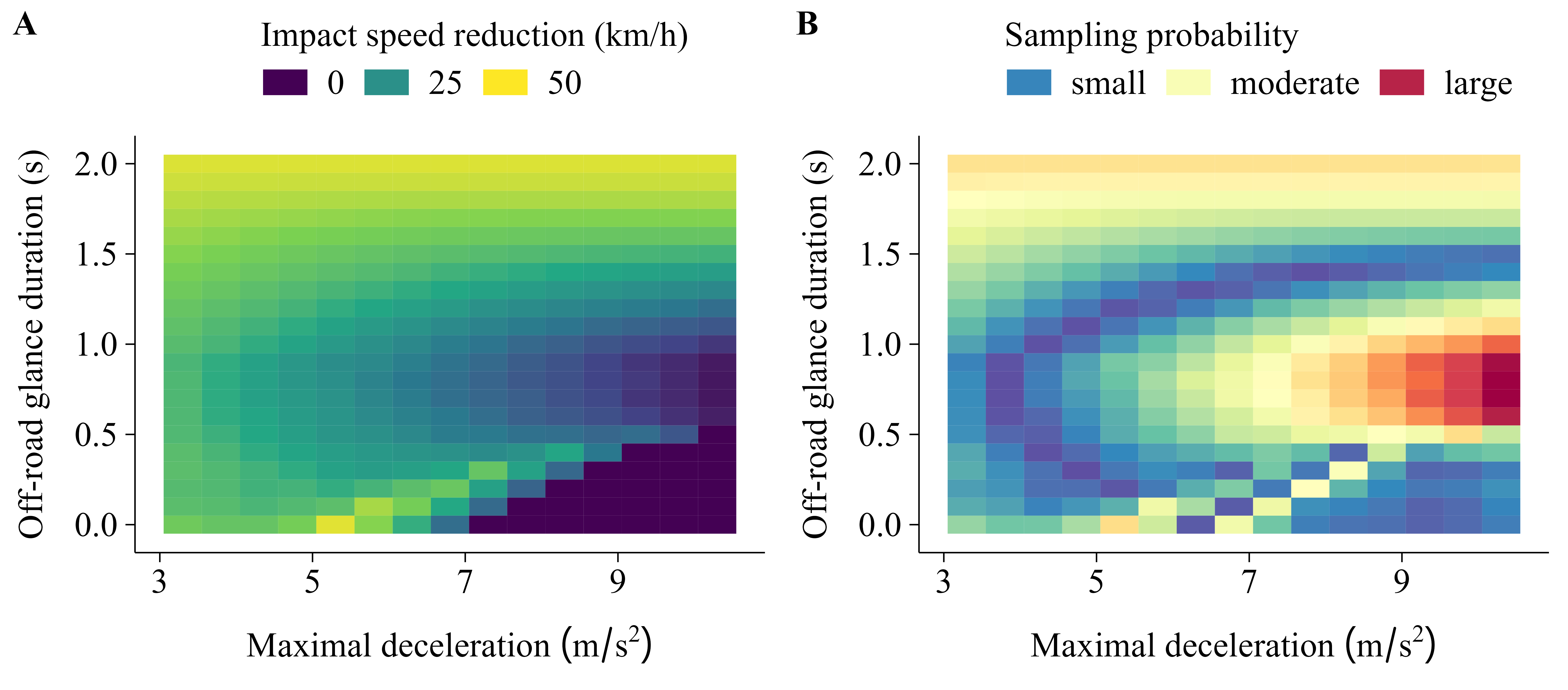}
    \caption{A: Simulated impact speed reduction with an automatic emergency braking system (AEB) compared to a baseline manual driving scenario (without AEB) in a computer experiment of a rear-end collision generation. In the bottom right corner, no crash was generated in the baseline scenario; such instances are non-informative with regards to safety benefit evaluation. B: Corresponding optimal active sampling scheme. Active sampling oversamples instances in regions where there is a high probability of generating a collision in the baseline scenario (attempting to generate only informative instances) and with a large predicted deviation from the average. These instances will be influential for estimating the safety benefit of the AEB system.}
    \label{fig:active_sampling_example}
\end{figure}

A small toy example of our problem and illustration of the proposed active sampling method is provided in Figure \ref{fig:active_sampling_example}. The figure shows the output of a computer simulation experiment to evaluate the impact speed reduction with an automatic emergency braking system (AEB) compared to a baseline manual driving scenario (without AEB) in a rear-end collision generation. The impact speed and impact speed reduction depend on the maximal deceleration during braking and the driver's off-road glance duration, i.e., the time the driver of the 'following' car is looking off-road  (e.g., due to distraction). By iteratively learning to predict the response surface of Figure \ref{fig:active_sampling_example} A while running the experiment, an accurate estimate of the overall safety benefit of the AEB system may be obtained by adaptive importance sampling (Figure \ref{fig:active_sampling_example} B). In doing so, computational demands can be substantially reduced compared to complete enumeration.

\section{Finite population sampling}
\label{sec:sampling}

We introduce the mathematical framework and notation in Section \ref{sec:mathematical_framwork}, presented in the context of the crash-causation-based scenario generation application outlined above. Traditional methods for sample selection and estimation are reviewed in Section \ref{sec:subsampling} and optimal importance sampling schemes discussed in Section \ref{sec:importance_sampling}.

\subsection{Target characteristic and scope of inference}
\label{sec:mathematical_framwork}

Assume we are given an index set or dataset $\mathcal D$ with $N$ instances or elements $i = 1, \ldots, N$. Associated with each element $i$ in $\mathcal D$ is a vector $(\boldsymbol{y}_i, \boldsymbol{z}_i)$, where $\boldsymbol{y}_i$ is a vector of outcomes or response variables, and $\boldsymbol{z}_i$ a vector of design variables and auxiliary variables. We are interested in a characteristic $\theta = h(\boldsymbol{t}_{\boldsymbol{y}})$ for some differentiable function $h:\mathbb{R}^d \to \mathbb{R}$ and $d$-dimensional vector of totals $\boldsymbol{t}_{\boldsymbol{y}} = \sum_{i=1}^N \boldsymbol{y}_i$.

In the context of crash-causation-based scenario generation, the index set $\mathcal D$ represents a collection of $N$ potential simulation scenarios of interest. The response variables $\boldsymbol{y}_i$ are outcomes of the simulation, including, e.g., whether a crash occurred or not, impact speed if there was a crash, and impact speed reduction with an advanced driver assistance system compared to some baseline driving scenario. The auxiliary variables $\boldsymbol{z}_i$ contain scenario information, such as simulation settings and parameters that are under the control of the investigator, and any additional information that is available without running the actual simulation. Characteristics of interest include, e.g., the mean impact speed reduction and crash avoidance rate with an advanced driver assistance system compared to some baseline driving scenario, when restricted to the relevant set of crashes (Figure \ref{fig:active_sampling_example}).

\subsection{Unequal probability sampling}
\label{sec:subsampling}

In our application, as in many computer simulation experiment applications,  running all $N$ simulations of interest to observe the outcomes $\{\boldsymbol{y}_i\}_{i=1}^N$ is computationally unfeasible. Hence, we assume that observing complete data is affordable only for a subset $\mathcal S \subset \mathcal D$ of size $n$. We consider the case when the subset $\mathcal S$ is selected using unequal probability sampling, i.e., by a random mechanism where each instance $i \in \mathcal D$ has a strictly positive and possibly unique probability of selection. In this section we also restrict ourselves to non-adaptive designs.  We let $S_{i}$ be the random variable representing the number of times an element $i$ is selected by the sampling mechanism, assuming that sampling may be with replacement. Hence, the subsample $\mathcal S$ is the random set given by $\mathcal S = \{i \in \mathcal D: S_i > 0\}$. We will primarily consider multinomial sampling designs but note that the methodology of our paper is applicable also for other designs, such as the Poisson sampling design \citep{Tille2006}, with minimal modifications.

In this context, an estimator for the finite population characteristic \eqref{eq:theta} may be obtained by sample weighting as
\begin{equation}
    \label{eq:hh}
    \hat{\theta} = h(\hat{\boldsymbol{t}}_{\boldsymbol{y}}), \quad 
    \hat{\boldsymbol{t}}_{\boldsymbol{y}} = \sum_{i \in \mathcal S} S_i w_i \boldsymbol{y}_i,
\end{equation}
where $w_i = 1/\mu_i$ and $\mu_i = \E[S_i]$. We note that $\hat{\boldsymbol{t}}_{\boldsymbol{y}}$ is an unbiased estimator of the total $\boldsymbol{t}_{\boldsymbol{y}}$ provided that $\mu_i > 0$ for all $i \in \mathcal D$, and furthermore a consistent estimator under general conditions \citep{Hansen1943, Horvitz1952}. Consequently, $\hat{\theta}$ is a consistent estimator for $\theta$ under mild assumptions \citep[see, e.g.,][]{Fuller2011}.

\subsection{Optimal importance sampling schemes}
\label{sec:importance_sampling}

When the function $h$ is linear and all $h(\boldsymbol{y}_i)$ are positive, it is well-known that the optimal sampling scheme for $\theta$ in terms of minimizing the variance of the estimator $\hat\theta$ is given by $\mu_i \propto h(\boldsymbol{y}_i)$, in fact producing an estimator with zero variance \citep{Fishman1996}. In general, one can show that the optimal importance sampling scheme for a characteristic $\theta = h(\boldsymbol{t}_{\boldsymbol{y}})$ and non-linear function $h(\boldsymbol{u})$ is of the form $\mu_i \propto \bigr\rvert\nabla h(\boldsymbol{\boldsymbol{t}_{\boldsymbol{y}}})^T\boldsymbol{y}_i\bigr\rvert$ (Proposition \ref{prop:naive_optimality}). A proof is provided in Appendix \ref{appendix:optimality}.

\begin{proposition}[Optimal importance sampling scheme, $\boldsymbol{y}_i$ known]
    \label{prop:naive_optimality}
    Let $\{\boldsymbol{y}_i\}_{i=1}^N$ be \\ fixed. Let $\{m_k\}_{k\ge 1}$ be an increasing sequence of positive integers and $\boldsymbol{S}_k = (S_{k1}, \ldots, S_{kN}) \sim \mathrm{Multinomial}(m_k, \boldsymbol{\pi})$ a corresponding sequence of random vectors. Let $\hat\theta_k$ be defined for the $k^{\mathrm{th}}$ random vector $\boldsymbol{S}_k$ as in \eqref{eq:hh}. As a function of $\boldsymbol{\pi} = (\pi_1, \ldots, \pi_N)$, the asymptotic mean squared error $\mathrm{AMSE}(\hat\theta) := \lim_{k\rightarrow \infty} \E[m_k(\hat\theta_k - \theta)^2]$ is minimized by 
    \begin{equation}
    \label{eq:optimalpi}
    \pi_i^* = \frac{\sqrt{c_i}}{\sum_{j=1}^N \sqrt{c_j}}, \quad i = 1, \ldots, N,
    \end{equation}
    with $c_i = \bigr\rvert\nabla h(\boldsymbol{u})^T\boldsymbol{y}_i\bigr\rvert^2_{\boldsymbol{u} = \boldsymbol{t}_{\boldsymbol{y}}}$.
\end{proposition}

We note that the result of Proposition \ref{prop:naive_optimality} is of limited practical use as it requires all the $\boldsymbol{y}_i$'s to be known.  Inspired by active learning \citep{Settles2012}, we introduce in Section \ref{sec:active_sampling_intro} an active sampling algorithm that overcomes this limitation through sequential sampling with iterative updates of the estimate for the total $\boldsymbol{t}_{\boldsymbol{y}}$ and predictions for the $\boldsymbol{y}_i$'s. However, as shown in the experiments in Section \ref{sec:experiments}, naively plugging in the predictions immediately to the importance sampling scheme of Proposition \ref{prop:naive_optimality} often results in poor performance. Indeed, accounting for prediction error is essential to control the variance of the active sampling estimator. We therefore in Proposition \ref{prop:practical_optimality} propose an optimal importance sampling scheme to minimize the expected mean squared error of our estimator for $\theta$, treating the unobserved values of the $\boldsymbol{y}_i$'s as random variables $\boldsymbol{Y}_i$. Integrated with flexible machine learning models, this will be the key ingredient of the active sampling method introduced in Section \ref{sec:active_sampling_intro}.

\begin{proposition}[Optimal importance sampling scheme, $\boldsymbol{y}_i$ unknown]
    \label{prop:practical_optimality}
    Let $\{\boldsymbol{y}_i\}_{i=1}^N$, $\boldsymbol{t}_{\boldsymbol{y}} = \sum_{i=1}^N \boldsymbol{y}_i$ and $\theta = h(\boldsymbol{t}_{\boldsymbol{y}})$ be fixed but unknown constants. Consider, as a proxy for $\boldsymbol{y}_i$, a collection of random variables $\{\boldsymbol {Y}_i\}_{i=1}^N$ with means $\E[\boldsymbol{Y}_i] = \boldsymbol{\eta}_i$ and finite, positive semi-definite covariance matrices $ \boldsymbol{\Cov}(\boldsymbol{Y}_i) = \boldsymbol{\Sigma}_i$. Let $m_k$, $\boldsymbol{S}_k$, $\hat{\theta}_k$ and $\mathrm{AMSE}(\hat\theta)$ be defined as in Proposition \ref{prop:naive_optimality}. Then, the expected asymptotic mean squared error $\E_{\boldsymbol{Y}}[\mathrm{AMSE}(\hat\theta)]$ is minimized by \eqref{eq:optimalpi} with $c_i = \left[(\nabla h(\boldsymbol{u})^T\boldsymbol{\eta}_i)^2 + \nabla h(\boldsymbol{u})^T \boldsymbol{\Sigma}_i \nabla h(\boldsymbol{u})\right]_{\boldsymbol{u} = \boldsymbol{t}_{\boldsymbol{y}}}$.
\end{proposition}

For a proof, see Appendix \ref{appendix:optimality}.

\section{Active sampling}
\label{sec:active_sampling_intro}

In this section we propose an active sampling strategy for finite population inference with optimal subsamples using adaptive importance sampling and machine learning. The active sampling algorithm is described in Section \ref{sec:active_sampling}. Variance estimation for the active sampling estimator is discussed in Section \ref{sec:variance_estimation} and asymptotic properties in Section 
\ref{sec:asymptotics}. We conclude by a brief discussion on sample size calculations for the active sampling method in Section \ref{sec:precision_calculation}.

\subsection{Active sampling algorithm}
\label{sec:active_sampling}
\FloatBarrier

The active sampling method is summarized in Algorithm \ref{alg:active_sampling}. The algorithm is executed iteratively in $K$ iterations $k = 1, \ldots, K$ and chooses, in each iteration, $n_k$ new instances at random (possibly with replacement) from the index set $\mathcal D = \{1, \ldots, N\}$. Once a new batch of instances has been selected we observe or retrieve the corresponding data $\boldsymbol{y}_i$ and update our estimates of the characteristics of interest. In our application, this is done by running a virtual computer simulation. The process continues until a pre-specified maximal number of iterations $K$ is reached, or the target characteristic is estimated with sufficient precision, based on a pre-specified precision target $\delta$ for the standard error of the estimator. Methods for variance estimation are discussed in Section \ref{sec:variance_estimation}.

\begin{algorithm}[htb!]
    \spacingset{1} 
    \caption{Active Sampling}
    \label{alg:active_sampling}
    \medskip

    \textbf{Input}: Index set $\mathcal D = \{1, \ldots, N\}$, target characteristic $\theta = h(\boldsymbol{t}_{\boldsymbol{y}})$ (to be estimated), precision target $\delta > 0$, maximal number of iterations $K$, batch sizes $\{n_k\}_{k = 1}^K$.
    
    \textbf{Initialization}: Let $m_0 = 0$, $\hat{\boldsymbol{t}}_{\boldsymbol{y}}^{(0)} = \boldsymbol{0}$, and $\mathcal{L}_0 = \varnothing$. 
     
    \begin{algorithmic}[1]
    \FOR{k = 1, 2, \ldots, K}

        \STATE \textbf{Learning} (only if $k > 1$): Train prediction model $f(\boldsymbol{y}_i|\boldsymbol{z}_i)$ on the labeled dataset $\{(\boldsymbol{y}_i, \boldsymbol{z}_i)\}_{i \in \mathcal L_{k-1}}$. Let $\hat{\boldsymbol{y}}_i$ and $\hat{\boldsymbol{\Sigma}}_i$ be the predicted mean and estimated residual covariance matrix for $\boldsymbol{Y}_i$, respectively. $^{*}$  
  	
        \IF {k $> 1$ and \textbf{Learning} step was successful$^{\dagger}$} 

            \STATE \textbf{Optimization}: Calculate sampling scheme $\boldsymbol{\pi}_k$ as
            \[
            \pi_{ki} \propto \sqrt{c_i}, 
            \quad 
            c_i = \left[(\nabla h(\boldsymbol{u})^T\hat{\boldsymbol{y}}_i)^2 + \nabla h(\boldsymbol{u})^T \hat{\boldsymbol{\Sigma}}_i \nabla h(\boldsymbol{u})\right]_{\boldsymbol{u} = \hat{\boldsymbol{t}}_{\boldsymbol{y}}^{(k-1)}}, \quad i \in \mathcal D.
            \]
        \ELSE
        
            \STATE \textbf{Fallback}: Set $\pi_{ki} \propto 1$ for all $i \in \mathcal D$.

        \ENDIF
        
        \STATE \textbf{Sampling}: Draw vector $\boldsymbol{s}_k = (s_{k1}, \ldots, s_{kN}) \sim \mathrm{Multinomial}(n_k, \boldsymbol{\pi}_k)$.
        
        \STATE \textbf{Labeling}: Retrieve data $\boldsymbol{y}_i$ for selected instance(s) $i : s_{ki} > 0$. Update labeled set $\mathcal{L}_k = \mathcal{L}_{k-1} \cup \{i \in \mathcal D: s_{ki} > 0\}$. 
		
       	\STATE \textbf{Estimation}: Let $\mu_{ki} = n_k \pi_{ki}$, $w_{ki} = 1/\mu_{ki}$, $m_k = m_{k-1} + n_k$, and 
        \begin{gather*}
            \hat{\boldsymbol{t}}_{\boldsymbol{y},k} = \sum_{i: s_{ki}>0} s_{ki}w_{ki}\boldsymbol{y}_i, \quad
            \hat{\boldsymbol{t}}_{\boldsymbol{y}}^{(k)} = \frac{1}{m_k}\left(m_{k-1}\hat{\boldsymbol{t}}_{\boldsymbol{y}}^{(k-1)} + n_k\hat{\boldsymbol{t}}_{\boldsymbol{y},k} \right), \quad
            \hat\theta^{(k)} = h(\hat{\boldsymbol{t}}_{\boldsymbol{y}}^{(k)}).
        \end{gather*}

        \STATE Estimate the variance of $\hat \theta^{(k)}$ according to \eqref{eq:deltamethod}. 

        \IF{$\sqrt{\widehat{\Var}(\hat \theta^{(k)})} < \delta$}
        
            \STATE \textbf{Termination}: Stop execution. Continue to \footnotesize{16}.

        \ENDIF

    \ENDFOR

    \STATE \textbf{Output}: Estimate $\hat\theta^{(k)}$, labeled dataset $\{(\boldsymbol{y}_i, \boldsymbol{z}_i)\}_{i \in \mathcal L_k}$ and selection history $\{\boldsymbol{s}_{j}, \boldsymbol{\mu}_{j}\}_{j=1}^{k}$. 
        
    \end{algorithmic}

    \medskip

    \footnotesize{$^{*}$Although the value of $\boldsymbol{y}_i$ is assumed to be fixed (but unknown) it is modeled here as a random variable $\boldsymbol{Y}_i$ to account for prediction uncertainty around the true value.}

    \footnotesize{$^{\dagger}$The prediction model could be fitted (converged and non-trivial model achieved) and prediction R-squared (regression) or prediction accuracy (classification) on hold-out data (e.g., by cross-validation) $>$ 0.}

    \medskip
\end{algorithm}

\spacingset{2}

A key component of the active sampling algorithm is the inclusion of an auxiliary model or surrogate model $f(\boldsymbol{y}|\boldsymbol{z})$ for the distribution of the unobserved data $\boldsymbol{y}_i$ given auxiliary variables $\boldsymbol{z}_i$. At this stage any prediction model or machine learning algorithm may be used. The first two moments of the response vector are then used as input to the optimal importance sampling scheme of Proposition \ref{prop:practical_optimality}. When the covariance matrices of the response vectors are not immediately available from the model, they may be estimated from the residuals. We suggest that this is done using the method of moments on hold-out data, e.g., by cross-validation. Underestimation of the residual variance may otherwise cause unstable performance by assigning sampling probabilities too close to zero with highly variable sample weights and increased estimation variance as a result. In practice, one may also need to make further simplifying assumptions, including assumptions about the mean-variance relationship and correlation structure of the response variables.

In each iteration $k$, the active sampling estimator $\hat{\theta}^{(k)}$ of the characteristic $\theta$ is constructed in three steps. First, we define an estimator $\hat{\boldsymbol{t}}_{\boldsymbol{y},k}$ for the total $\boldsymbol{t}_{\boldsymbol{y}}$ using data acquired in the current iteration. This estimator is then combined with the estimators from the previous iterations to produce a pooled estimator $\hat{\boldsymbol{t}}_{\boldsymbol{y}}^{(k)}$. Finally, our estimator for $\theta$ is obtained using the plug-in estimator $h(\hat{\boldsymbol{t}}_{\boldsymbol{y}}^{(k)})$. We note that the pooled estimator $\hat{\boldsymbol{t}}_{\boldsymbol{y}}^{(k)}$ is an unbiased estimator for the finite population total, provided that $\pi_{ki}>0$ for all $k$ and $i$. Consequently, one may expect our estimator $\hat\theta^{(k)}$ to be consistent for $\theta$ under mild assumptions. We will return to this in Section \ref{sec:asymptotics}.

By gathering data in a sequential manner, we are able to learn from past observations how to sample in an optimal way in future iterations. The proposed active sampling scheme interpolates in a completely data-driven manner between simple random sampling when the prediction error is large (or no model has been fitted) and the optimal importance sampling scheme of Proposition \ref{prop:naive_optimality} when the prediction error is small. Importantly, unbiased inferences for $\theta$ are obtained even if the surrogate model $f(\boldsymbol{y}|\boldsymbol{z})$ would be biased. This is due to the use of importance sampling and inverse probability weighting. However, the performance of the active sampling algorithm in terms of variance depends on the adequacy of the prediction model and capability of capturing the true signals in the data. It also depends on the signal-to-noise-ratio between the inputs or auxiliary variables $\boldsymbol{z}_i$ and response vectors $\boldsymbol{y}_i$. The stronger the association, the greater the potential benefit of active sampling.

\subsection{Variance estimation}
\label{sec:variance_estimation}

To estimate the variance of our estimator $\hat\theta^{(k)}$, we first need an estimator of the covariance matrix $\boldsymbol{\Psi}^{(k)} = \COV(\hat{\boldsymbol{t}}_{\boldsymbol{y}}^{(k)})$ for the pooled estimator $\hat{\boldsymbol{t}}_{\boldsymbol{y}}^{(k)}$ of the finite population total $\boldsymbol{t}_{\boldsymbol{y}}$. Given such an estimate, the variance of $\hat{\theta}^{(k)} = h(\hat{\boldsymbol{t}}_{\boldsymbol{y}}^{(k)})$ may be estimated using the delta method as
\begin{equation}
    \label{eq:deltamethod}
    \widehat{\Var}(\hat \theta^{(k)}) = \nabla h(\boldsymbol{u})^T\hat{\boldsymbol{\Psi}}^{(k)} \nabla h(\boldsymbol{u})\bigr\rvert_{\boldsymbol{u} = \hat{\boldsymbol{t}}_{\boldsymbol{y}}^{(k)}},   
\end{equation}
\citep[see, e.g.,][]{Sen1993}. Three different approaches to variance estimation are presented below and evaluated empirically in Section \ref{sec:application}. A theoretical justification is provided by  Proposition \ref{prop:martingaleCLT} and Corollary \ref{cor:multivariateMartingaleCLT} in Appendix \ref{appendix:clt}. 

\paragraph{Method 1 (Design-based variance estimator):} First, we may proceed as for the estimator of the finite population total $\boldsymbol{t}_{\boldsymbol{y}}$ and use a pooled variance estimator    
\begin{align*}
    \hat{\boldsymbol{\Psi}}_1^{(k)} & = m_k^{-2}\sum_{j=1}^k n_j^2 \hat{\boldsymbol{\Phi}}_{j},
\end{align*}    
where $\hat{\boldsymbol{\Phi}}_{j}$ are (any) unbiased estimators of the conditional covariance matrices $\boldsymbol{\Phi}_{j}= \COV(\hat{\boldsymbol{t}}_{\boldsymbol{y},j}|\boldsymbol{S}_1, \ldots, \boldsymbol{S}_{j-1})$. Each of the covariance matrices $\boldsymbol{\Phi}_{j}$ may be estimated using standard survey sampling techniques. For instance, under the multinomial design we may use Sen-Yates-Grundy estimator for $\boldsymbol{\Phi}_j$, i.e.,
\begin{equation*}
    \hat{\boldsymbol{\Phi}}_{j} = \frac{n_j}{n_j - 1} \sum_{i \in \mathcal D} S_{ji}\left(\frac{\boldsymbol{y}_i}{\mu_{ji}} - \frac{\hat{\boldsymbol{t}}_{\boldsymbol{y},j}}{n_j}\right)\left(\frac{\boldsymbol{y}_i}{\mu_{ji}} - \frac{\hat{\boldsymbol{t}}_{\boldsymbol{y},j}}{n_j}\right)^T, \quad \mu_{ji} = n_j\pi_{ji},
\end{equation*}
provided that $n_j \ge 2$ \citep{Sen1953, Yates1953}. For fixed-size designs with $n_j=1$, other estimators must be used.

\paragraph{Method 2 (Martingale variance estimator):} Alternatively, we may use the squared variation of the individual estimates $\hat{\boldsymbol{t}}_{\boldsymbol{y},j}$ and estimate $\boldsymbol{\Psi}^{(k)}$ by
\begin{equation*}
    \hat{\boldsymbol{\Psi}}_2^{(k)} = m_k^{-2}\sum_{j=1}^k n_j^2\left(\hat{\boldsymbol{t}}_{\boldsymbol{y},j}-\hat{\boldsymbol{t}}_{\boldsymbol{y}}^{(k)}\right)\left(\hat{\boldsymbol{t}}_{\boldsymbol{y},j}-\hat{\boldsymbol{t}}_{\boldsymbol{y}}^{(k)}\right)^T. 
\end{equation*}
This estimator arises immediately from the martingale theory used for our asymptotic analyses in Appendix \ref{appendix:clt}. This method is particularly useful when the batch sizes are small and the number of iterations large. 

\paragraph{Method 3 (Bootstrap variance estimator):} Finally, variance estimation may be conducted by non-parametric bootstrap \citep{Efron1979, Davison1997}. If subsampling is done with replacement, the importance-weighted bootstrap should be used to account for possible differences in the number of selections per observation. Specifically, the bootstrap sample size should be equal to the total sample size $m_k = \sum_{j=1}^k n_j$ (number of distinct selections), and the selection probabilities for the bootstrap proportional to the number of selections $\sum_{j=1}^k s_{ji}$ per instance $i$. One way to achieve this with ordinary bootstrap software is to create an augmented dataset with one record for each of the $s_{ji}$ selections, and perform ordinary non-parametric bootstrap on the augmented dataset. An estimate of the covariance matrix of $\hat{\boldsymbol{t}}_{\boldsymbol{y}}^{(k)}$ can then be obtained by
\begin{equation*}
    \hat{\boldsymbol{\Psi}}_3^{(k)} = \frac{1}{B-1}\sum_{b=1}^B \left(\tilde{\boldsymbol{t}}_{\boldsymbol{y},b}^{(k)}-\bar{\boldsymbol{t}}_{\boldsymbol{y}}^{(k)}\right)\left(\hat{\boldsymbol{t}}_{\boldsymbol{y},b}^{(k)}-\bar{\boldsymbol{t}}_{\boldsymbol{y}}^{(k)}\right)^T, 
\end{equation*}
where $\bar{\boldsymbol{t}}_{\boldsymbol{y}}^{(k)} = \frac{1}{B}\sum_{b=1}^B \tilde{\boldsymbol{t}}_{\boldsymbol{y},b}^{(k)}$ is the mean of $B$ bootstrap estimates $\tilde{\boldsymbol{t}}_{\boldsymbol{y},b}^{(k)}$ of $\boldsymbol{t}_{\boldsymbol{y}}$.

\subsection{Asymptotic properties and interval estimation}
\label{sec:asymptotics}

Using the martingale central limit theorem of \citet{Brown1971}, we show that under mild assumptions our active sampling estimator is consistent and asymptotically normally distributed, for fixed $N$ and bounded batch sizes $n_k$, as the number of iterations tends to infinity (Proposition \ref{prop:martingaleCLT} and Corollary \ref{cor:multivariateMartingaleCLT}, Appendix \ref{appendix:clt}). The essential conditions for this to hold are (in the scalar case) that:
\begin{enumerate}[label=\roman*)]
    \item the sampling probabilities are properly bounded away from zero,  
    \item the total variance $\Var(\sum_{j=1}^k \hat t_{y,j})$ tends to infinity as the number of iterations $k \rightarrow \infty$, and 
    \item     the ratio of the total variance $\Var(\sum_{j=1}^k \hat t_{y,j})$ to the sum of conditional variances \\$\sum_{j=1}^k\Var(\hat t_{y,j}|\boldsymbol{S}_1, \ldots, \boldsymbol{S}_{j-1})$ converges in probability to $1$  as $k \rightarrow \infty$.  
\end{enumerate}
Similar conditions are sufficient also for consistent variance estimation. In this setting, we note that the importance sampling scheme in Algorithm \ref{alg:active_sampling} remains optimal in the sense of minimizing the variance contribution (or mean squared error contribution) from each iteration of the algorithm, given the information available so far.

In practice, the first assumption may be violated by overfitting and underestimation of the residual variance in the learning step of the active sampling algorithm. Both of these issues may cause variance inflation and an erratic behavior of the estimator due to incidentally large sample weights. The second assumption could be violated, e.g., for a linear estimator in a noise-free setting where a perfect importance sampling scheme yielding zero variance may be found. Indeed, an optimal importance sampling estimator would in this case have zero variance and hence would not converge towards a normal limit. In most cases, however, estimation- and prediction uncertainty are intrinsic to the problem, and the second assumption is trivially fulfilled in most realistic applications. The third assumption is more of technical nature and needed to ensure that the statistical properties of the active sampling estimator can be deduced from a single execution of the algorithm. Empirical justification for these assumptions is provided in Section \ref{sec:application}.

Confidence intervals can be calculated using the classical large sample formula
\begin{equation}
    \label{eq:confidence_interval}
    \hat\theta^{(k)} \pm z_{\alpha/2} \times \SE_{\hat \theta^{(k)}} 
\end{equation}
where $\hat \theta^{(k)}$ is the estimate of the characteristic $\theta$, $\SE_{\hat \theta^{(k)}} = \sqrt{\widehat{\Var}(\hat \theta^{(k)})}$ the corresponding standard error, and $z_{\alpha/2}$ the $\alpha/2$-quantile of a standard normal distribution. Under the assumptions stated above, such a confidence interval has approximately $100 \times (1 - \alpha)$\% coverage of the true population characteristic $\theta$, under repeated subsampling from $\mathcal D$, in large enough samples.

\subsection{How many samples are needed?}
\label{sec:precision_calculation}

An important practical question is how many samples or iterations of the active sampling algorithm that are required for estimating a characteristic $\theta$ with sufficient precision. This question can be addressed as follows. First, a pilot sample may be selected to obtain an initial estimate of $\theta$ with a corresponding estimate for the variance. A precision calculation may then be conducted using standard theory for simple random sampling designs, and the number of samples needed for a certain level of precision deduced. This would give a conservative estimate of the sample size needed for the active sampling algorithm, which usually can be terminated for sufficient precision with much smaller samples. Importantly, the pilot sample can be re-used in the first iteration of the active sampling algorithm and hence comes at no additional cost. It also possible to monitor the precision of the active sampling estimator during execution of the algorithm and possibly update the precision target or iteration limit as needed.

\FloatBarrier

\section{Simulation experiments}
\label{sec:experiments}

We evaluated the empirical performance of the active sampling method by repeated subsampling on synthetic data. Methods are described in Section \ref{sec:experiments_methods} and results in Section \ref{sec:experiments_results}.

\subsection{Data and Methods}
\label{sec:experiments_methods}

\begin{figure}[htb!]
    \centering
    \includegraphics[scale = 1]{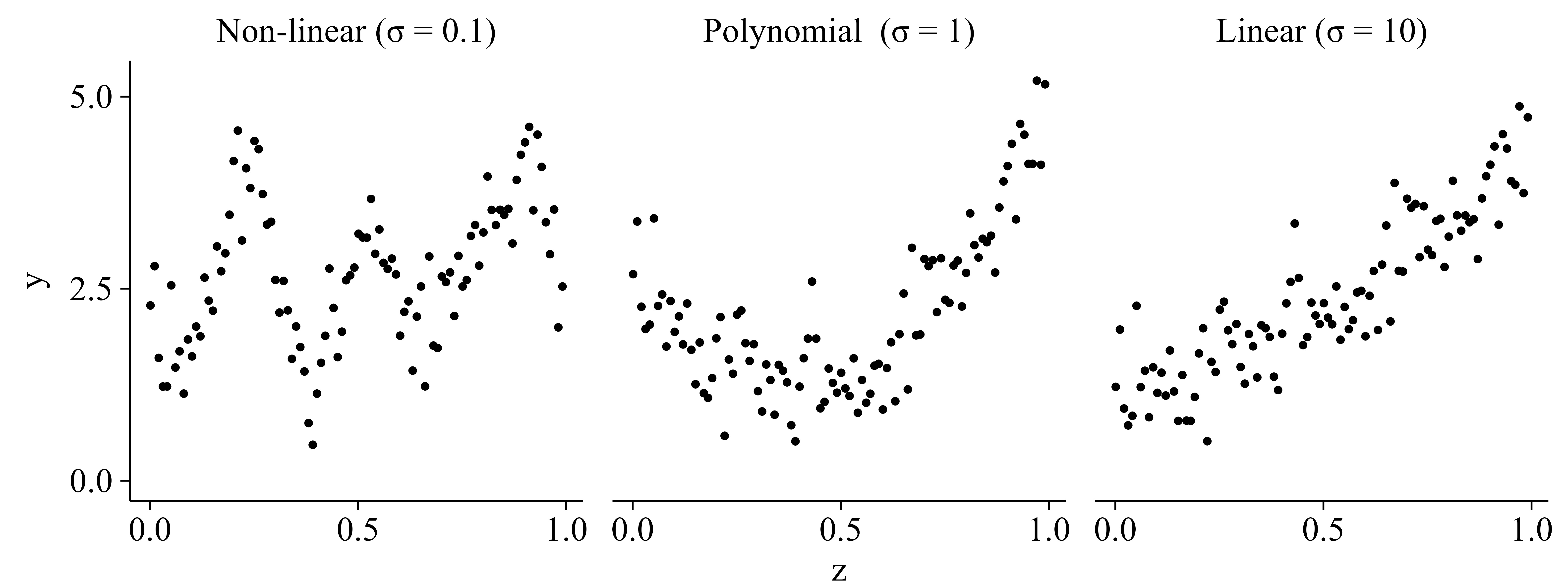}
    \caption{Examples of three synthetic datasets with varying degree of non-linearity. Data were generated according to a Gaussian process, using a Gaussian kernel with bandwidth $\sigma$.}
    \label{fig:nonlin}
\end{figure}

We generated a total of 24 datasets with varying support, signal-to-noise-ratio, and degree of non-linearity in the association between a scalar auxiliary variable $z_i$ and scalar response variable $y_i$. This was done as follows. First, $N=10^3$ data points $z_i$ were generated on a uniform grid from $0.001$ to $1$. This was taken as our auxiliary variable. Next we generated a variable $y_i$ according to a Gaussian process, using a Gaussian kernel with bandwidth $\sigma$. This was taken as the study variable of interest. We varied the bandwidth $\sigma = 0.1, 1, 10$, corresponding to a non-linear, polynomial, and linear scenario (Figure \ref{fig:nonlin}). We also varied the residual variance to obtain a coefficient of determination $R^2 = 0.10, 0.50, 0.75, 0.90$ for the true model, corresponding to a low, moderate, high, and very high signal-to-noise ratio. Finally, we normalized the response variable to have unit variance, positive correlation with the auxiliary variable, and support on the positive real line (strictly positive scenario, $\min_{1\le i \le N} y_i = 0.1$) or zero mean (unrestricted scenario, $\bar y = 0$, $y_i \in \mathbb{R}$).

We used active sampling to estimate the finite population mean $\bar y$ using a linear estimator $h(u) = u / N, \boldsymbol{y}_i = y_i$, and non-linear (H{\'a}jek) estimator $h(\boldsymbol{u}) = u_2 / u_1, \boldsymbol{y} = (1, y_i)^T$.\footnote{In the non-adaptive setting, the linear estimator is given by $N^{-1}\sum_{i\in \mathcal S} S_iw_iy_i$ and the H{\'a}jek estimator by $\hat N^{-1}\sum_{i \in \mathcal S} S_iw_iy_i$, $\hat N = \sum_{i \in \mathcal S}S_iw_i$.} The active sampling algorithm was implemented according to Algorithm \ref{alg:active_sampling} with a batch size of $n_k = 10$ or $n_k = 50$ observations per iteration. The learning step was implemented using a simple linear regression model, generalized additive model (thin plate spline), random forests, gradient boosting trees, and Gaussian process regression surrogate model for $y_i$ given $z_i$. For comparison we implemented simple random sampling using the before-mentioned estimators (linear and non-linear), control variate estimator \citep{Fishman1996}, and ratio estimator \citep{Särndal2003}. We also compared to importance sampling with probability proportional to the auxiliary variable $z_i$. We finally implemented a naive version of the active sampling algorithm ignoring prediction uncertainty, i.e., setting the residual covariance matrix equal to zero in the optimization step of the algorithm. This is the same as to plug in the predictions from the surrogate model into the formula for the theoretically optimal sampling scheme of Proposition \ref{prop:naive_optimality}, treating the predictions as known true values of the $y_i$'s. Each sampling method was repeated 500 times for sample sizes up to $n = 250$ observations.

The performance was measured by the root mean squared error of the estimator (eRMSE) for the finite population mean $\bar y$, calculated as
\begin{equation}
    \label{eq:rmse}
    \mathrm{RMSE}(\hat{\theta}) = \sqrt{\frac{1}{m}\sum_{m=1}^M (\hat\theta_{m}(n) - \theta)^2},
\end{equation}
where $\hat\theta_{m}(n)$ is the estimate in the $m^{\mathrm{th}}$ simulation from a sample of size $n$, $M=500$ the number of simulations, and $\theta = \bar y$ the characteristic of interest (i.e., the ground truth).

The experiments were implemented using the \texttt{R} language and environment for statistical computing \citep{R_core_team}, version 4.2.3. The \texttt{R} code is available online at \href{https://github.com/imbhe/ActiveSampling}{\nolinkurl{https://github.com/imbhe/ActiveSampling}}.

\subsection{Results}
\label{sec:experiments_results}

\begin{figure}[htb!]
    \centering
    \includegraphics[scale = 1]{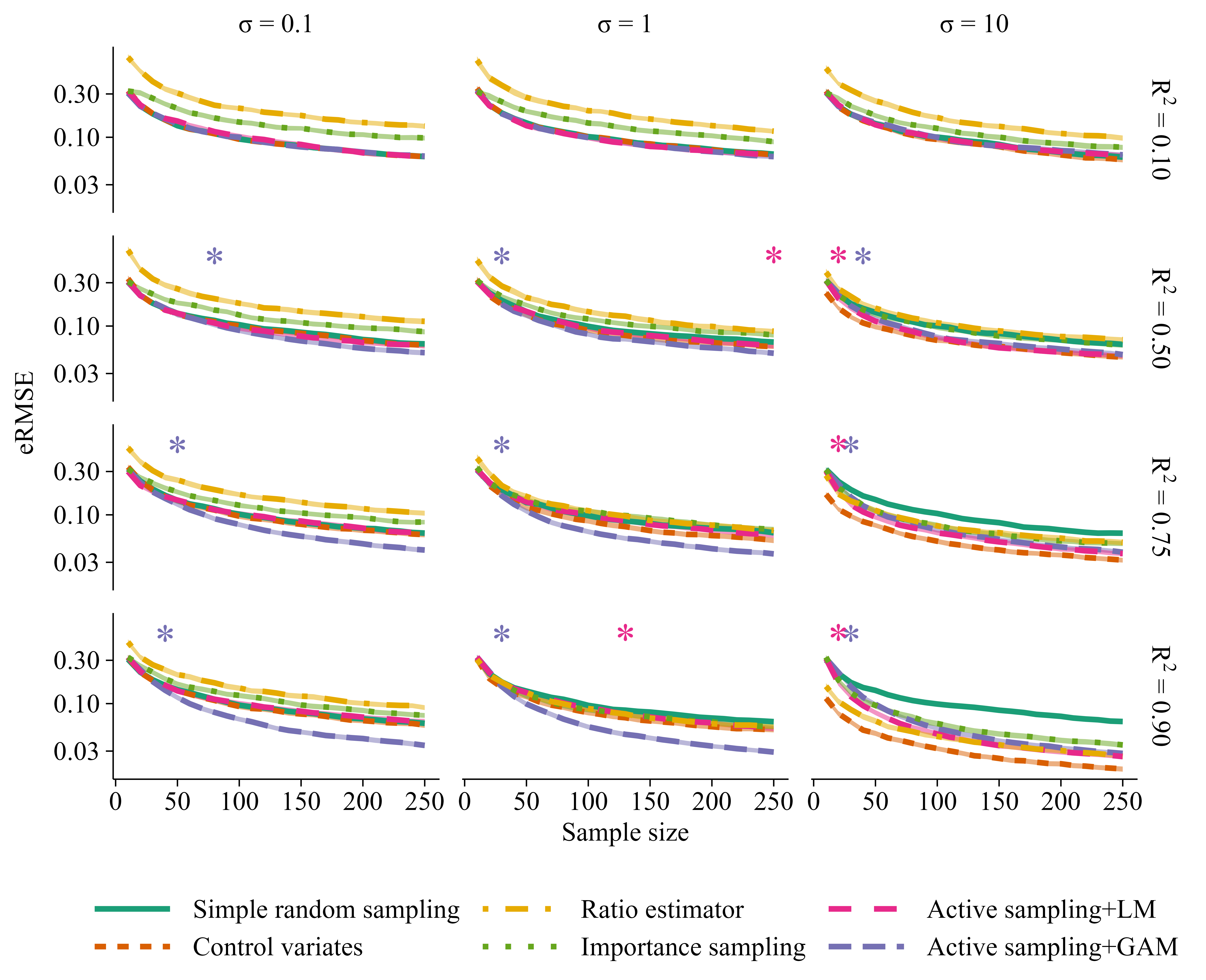}
    \caption{Performance of active sampling using a linear surrogate model (LM) or generalized additive surrogate model (GAM) compared to simple random sampling, ratio estimator, control variates, and importance sampling for estimating a finite population mean in a strictly positive scenario (all $y_i>0$) using a linear estimator ($h(u) = u / N$) and batch size $n_k=10$. Results are shown for 12 different scenarios with varying signal-to-noise ratio ($R^2$) and varying degree of non-linearity ($\sigma$) (cf. Figure \ref{fig:nonlin}). Shaded regions are 95\% confidence intervals for the root mean squared error of the estimator (eRMSE) based on 500 repeated subsampling experiments. Asterisks show the smallest sample sizes for which there were persistent significant differences ($p<.05$) between active sampling and simple random sampling.}
    \label{fig:simulation_experiments_main_result}
\end{figure}

The results of active sampling compared to four benchmark methods are shown in Figure \ref{fig:simulation_experiments_main_result} for the strictly positive scenario, linear estimator, batch size $n_k=10$, and linear or generalized additive surrogate model. Results under other settings are presented in Appendix \ref{appendix:supplemental_figures_experiments}.

There were substantial reductions in eRMSE with active sampling compared to both simple random sampling and standard variance reduction techniques in the non-linear ($\sigma=0.1$) and polynomial ($\sigma=1$) scenarios when a generalized additive surrogate model was used (Figure \ref{fig:simulation_experiments_main_result}). Similar results were observed also using random forests, gradient boosting trees, and Gaussian process regression as surrogate models (Figure \ref{fig:all_ml_models}, Appendix \ref{appendix:supplemental_figures_experiments}). In contrast, there was a slight advantage of the standard variance reduction techniques in the linear setting ($\sigma=10$). Batch size influenced the performance, with a better performance when using a smaller ($n_k=10$) compared to larger batch size ($n_k=50$). However, the effect of batch size was attenuated as the number of iterations increased (Figure \ref{fig:batch_size}, Appendix \ref{appendix:supplemental_figures_experiments}). The benefits of active sampling were somewhat smaller in the unrestricted scenario ($\bar y = 0$, $y_i \in \mathbb{R}$) and for non-linear estimators. Still, sample size reductions of up to 30\% were achieved compared to simple random sampling for the same level of performance (Figure \ref{fig:zero_mean} and \ref{fig:hajek}, Appendix \ref{appendix:supplemental_figures_experiments}).

Notably, active sampling never performed worse than simple random sampling, even for a misspecified model (i.e., when applying a linear surrogate model to non-linear data; Figure \ref{fig:simulation_experiments_main_result}, Figure \ref{fig:zero_mean} and \ref{fig:hajek} in Appendix \ref{appendix:supplemental_figures_experiments}). In contrast, a naive implementation of the active sampling algorithm, ignoring prediction uncertainty, resulted in worse performance than simple random sampling. This was particularly exacerbated in low signal-to-noise ratio settings, for non-positive data, non-linear estimators, and misspecified models (Figure \ref{fig:naive_strictly_positive} and \ref{fig:naive_hajek}, Appendix \ref{appendix:supplemental_figures_experiments}).

\section{Application}
\label{sec:application}

We next implemented active sampling on the crash-causation-based scenario generation problem introduced in Section \ref{sec:intro_application}. The data, model, and simulation set-up is described in Section \ref{sec:data and methos}, together with methods for performance evaluation. Empirical results are presented in Section \ref{sec:results}.

\subsection{Data and Methods}
\label{sec:data and methos}

\paragraph{Ground truth dataset} 
The data used for scenario generation in this study were reconstructed pre-crash kinematics of 44 rear-end crashes from a crash database provided by Volvo Car Corporation. This database contains information about crashes that occurred with Volvo vehicles in Sweden \citep{isaksson2005thirty}. We constructed a ground truth dataset by running virtual simulations for all 1,005 combinations of glance duration (67 levels, 0.0–6.6s) and deceleration (15 levels, 3.3–10.3 m/s$^2$) for all 44 crashes. Additionally, each scenario configuration was associated with a prior probability $p_i$ of occurring in real life, estimated by the empirical probability distribution of the glance-deceleration distribution in real crashes. The simulations were run under both manual driving (baseline scenario) and automated emergency braking (AEB) system conditions, producing a dataset of 44,220 pairs of observations. Running the complete set of simulations took about 50 hours, running 26 threads in parallel on a high-performance computer equipped with 24 Intel$^{\circledR}$ Xeon$^{\circledR}$ CPU E5-2620 processors.

\paragraph{Outcomes and measurements} 
The outputs of the simulations were the impact speed under both scenarios (baseline and AEB). We also calculated the impact speed reduction (continuous) and crash avoidance (binary) of the AEB system compared to the baseline scenario. The aim in our experiments was to estimate the benefit of the AEB system, as measured by mean impact speed reduction and crash avoidance rate compared to baseline manual driving, given that there was a crash in the baseline scenario. Accounting for the prior observation weights (scenario probabilities) $p_i$, the target characteristic $\theta$ may in this case be written as
\begin{equation}
    \label{eq:theta_application}
    \theta = \frac{\sum_{i=1}^N p_i (y_{i,0} - y_{i,1}) I(y_{i,0}>0)}{\sum_{i=1}^N p_i I(y_{i,0}>0)} 
\end{equation}
where $y_{i,0}$ is the outcome of the simulation (e.g, impact speed or binary crash indicator) under the baseline scenario, $y_{i,1}$ the corresponding outcome with the countermeasure (AEB), and $I(y_{i,0}>0)$ a binary indicator taking the value $1$ if there was a collision in the baseline scenario and $0$ otherwise. The observation weights $p_i$ are known \textit{a priori} and need not be learned from data. This makes our problem particularly suitable for importance sampling methods. Note also that there may be large regions in the input space generating no crash (cf. Figure 1), hence providing no information for the characteristic $\theta$. Active sampling offers an opportunity to learn and exploit this feature during the sampling process.

As auxiliary variables we used the glance duration and maximal deceleration during braking, i.e., the inputs to the virtual simulation experiment, and an \textit{a priori} known maximal impact speed per original crash event. The maximal impact speed was considered as a means to summarize a 44-level categorical variable (ID of the original crash event) as a single numeric variable in the random forest algorithm used for the learning step of the active sampling method; see \textbf{Implementation} below and Appendix \ref{appendix:methods} for further details. Although comprising only three variables, this corresponds with ordinary statistical methods to an 88-dimensional vector of auxiliary variables (or greater, if non-linear terms are included), counting all the interactions between glance duration and deceleration with the 44 original crash events.

\paragraph{Confidence interval coverage rates}
We evaluated the large-sample normal confidence intervals \eqref{eq:confidence_interval} with the three different methods for variance estimation described in Section \ref{sec:variance_estimation}: the design-based (pooled Sen-Yates-Grundy) estimator, martingale estimator, and bootstrap estimator. The empirical coverage rates of the confidence intervals were calculated using 500 repeated subsampling experiments.

\paragraph{Active sampling performance evaluation}
We evaluated the performance of the active sampling method for estimating the mean impact speed reduction or crash avoidance rate of an AEB system compared to baseline driving (without AEB). Active sampling performance was evaluated against simple random sampling, importance sampling, Latin hypercube sampling \citep{Cioppa2007, Meng2021}, leverage sampling \citep{Ma2015, Ma2022}, and active learning with Gaussian processes. Two importance sampling schemes were considered: a density sampling scheme with probabilities proportional to the prior observation weights $p_i$, and a severity sampling scheme that additionally attempts to oversample high-severity instances (i.e., with low deceleration and long glances). Each subsampling method was repeated 500 times up to a total sample size of $n = 2,000$ observations. The performance was measured by the root mean squared error of the estimator (eRMSE) compared to ground truth, calculated as in \eqref{eq:rmse}. The results are presented graphically as functions of the sample size, i.e., the number of baseline-AEB simulations pairs.

\paragraph{Implementation}
\label{para:implementation}
The empirical evaluation was implemented using the \texttt{R} language and environment for statistical computing, version 4.2.1 \citep{R_core_team}. Active sampling was implemented with a batch size of $n_k=10$ observations per iteration. Random forests \citep{Breiman2001} were used for the learning step of the algorithm. We also performed sensitivity analyses for the choice of machine learning algorithm using extreme gradient boosting \citep{Chen2016} and k-nearest neighbors. Latin hypercube sampling was implemented similarly to \cite{Meng2021}. Statistical leverage scores for the leverage sampling method were calculated using weighted least squares with the two auxiliary variables (off-road glance duration and maximal deceleration during braking) as explanatory variables and the prior scenario probabilities $p_i$ as weights. The Gaussian process active learning method was implemented using a probabilistic uncertainty scheme, with probabilities proportional to the posterior uncertainty (standard deviation) of the predictions. This was chosen based on computational considerations and to promote exploration of the design space. For Gaussian process active learning, estimation was conducted using a model-based estimator by evaluating the predictions over the entire input space. All other methods used observed data rather than predicted values for estimation. Further implementation details are provided in Appendix \ref{appendix:methods}. The \texttt{R} code and data are available online at \href{https://github.com/imbhe/ActiveSampling}{\nolinkurl{https://github.com/imbhe/ActiveSampling}}.

\subsection{Results}
\label{sec:results}

\paragraph{Confidence interval coverage rates}
\label{sec:CI_coverage}

\begin{figure}[htb!]
    \begin{center}
        \includegraphics[scale = 1]{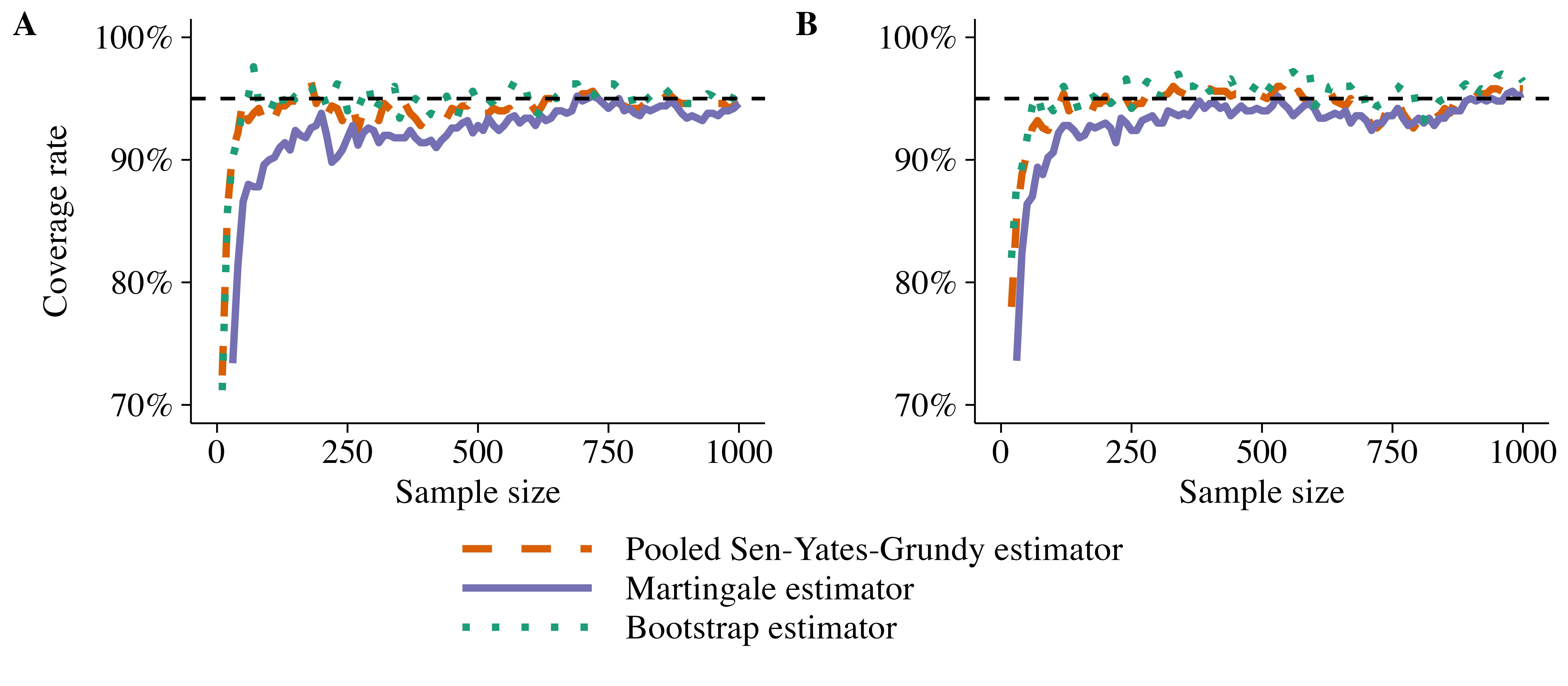}
    \end{center}
    \caption{Empirical coverage rates of 95\% confidence intervals for the mean impact speed reduction (A) and crash avoidance rate (B) using active sampling. The lines show the coverage rates with three different methods for variance estimation in 500 repeated subsampling experiments. A batch size of $n_k=10$ observations per iteration was used.}
    \label{fig:95_coverage_plot}
\end{figure}

The empirical coverage rates of large-sample normal confidence intervals under active sampling are presented in Figure \ref{fig:95_coverage_plot}. There was a clear under-coverage in small samples, as expected. Both the pooled Sen-Yates-Grundy estimator and bootstrap variance estimator produced confidence intervals that approached the nominal 95\% confidence level relatively quickly as the sample size increased. Coverage rates were somewhat lower with the martingale variance estimator, and more iterations where needed before the nominal 95\% level was reached.

\begin{figure}[htb!]
    \begin{center}
    \includegraphics[scale = 1]{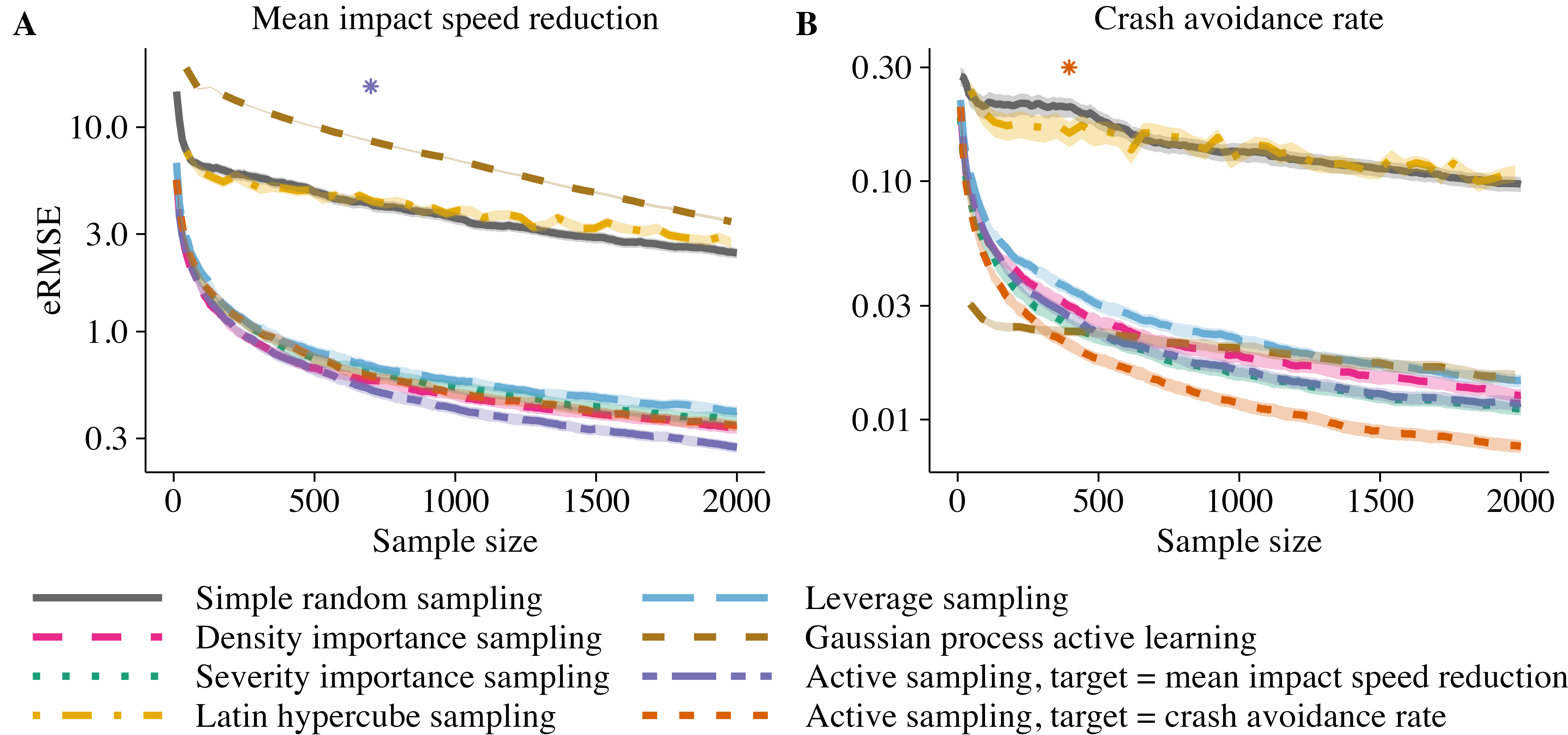}
    \end{center}
    \caption{Root mean squared error (eRMSE) for estimating the mean impact speed reduction (A) and crash avoidance rate (B). The lines show the performance using simple random sampling, importance sampling, Latin hypercube sampling, leverage sampling, Gaussian process active learning, and active sampling optimized for the estimation of the mean impact speed reduction and crash avoidance rate. Shaded regions represent 95\% confidence intervals for the eRMSE based on 500 repeated subsampling experiments. Asterisks show the smallest sample sizes for which there were persistent significant differences ($p<.05$) between active sampling and the best performing benchmark method.}
    \label{fig:active_sampling_vs_importance_sampling}
\end{figure}

\paragraph{Active sampling performance evaluation}
The eRMSE with active sampling compared to five benchmark methods is presented in Figure \ref{fig:active_sampling_vs_importance_sampling}. Simple random sampling and Latin hypercube sampling overall performed worst and had similar performance. Active learning using Gaussian processes had good performance for the crash avoidance (which was relatively constant over the input space, with ~80\% of all crashes avoided by the AEB), but poor performance for the impact speed reduction (which varied more and was harder to predict). With the other methods, estimation variance in the early iterations was largely driven by the variance of the scenario probability weights in the subsample. In contrast, estimation variance for the model-based (Gaussian process) response surface estimator was attenuated by evaluating predictions over the entire input space. Leverage sampling and the two importance sampling schemes had similar performance, with a slight advantage of severity importance sampling for estimating the crash avoidance rate. Active sampling optimized for a specific characteristic had best performance on the characteristic for which it was optimized. Significant improvements compared to the best performing benchmark method were observed from around 400 samples for estimating the crash avoidance rate and 700 observations for estimating the mean impact speed reduction.

The benefit of active sampling increased with the sample size. At $n=2,000$ observations, a reduction in eRMSE of 20–39\% was observed compared to importance sampling. Accordingly, active sampling required up to 46\% fewer observations than importance sampling to reach the same level of performance on the characteristic for which it was optimized. Moreover, active sampling performance was on par with that of traditional methods when evaluated on characteristics other than the one it was optimized for. Similar results were observed when using k-nearest neighbors and extreme gradient boosting as auxiliary models for the learning step of the active sampling algorithm (Figure \ref{fig:different_prediction_models_active_sampling_vs_importance_sampling}, Appendix \ref{appendix:supplemental_figures_application}).

Active sampling was also relatively fast and required about 60 seconds for running $200$ iterations (generating $n=2,000$ samples) on a laptop computer equipped with an AMD Ryzen 7 PRO 585OU 1.90 GHz processor. The Gaussian process active learning method required approximately 270 seconds to generate the same number of samples.

\section{Discussion}
\label{sec:discussion}

We have presented an active sampling framework for finite population inference with optimal subsamples. Active sampling outperformed standard variance reduction techniques in non-linear settings, and also in linear settings with moderate signal-to-noise ratio. We evaluated the performance of active sampling for safety assessment of advanced driver assistance systems in the context of crash-causation-based scenario generation. Substantial improvements over traditional importance sampling methods were demonstrated, with sample size reductions of up to 50\% for the same level of performance in terms of eRMSE. In our application, active sampling was also superior to space-filling, leverage sampling, and Gaussian process active learning methods.

Our work contributes to the ongoing development of subsampling methods in statistics and for computer simulation experiments in particular. In this context, Gaussian processes and space-filling methods have been particularly popular and shown great success for a variety of tasks \citep{Cioppa2007, Sun2017, feng2020testing, batsch2021scenario, Lim2021}. In our application, however, neither space-filling methods nor Gaussian process active learning performed as well as importance sampling or active sampling. Although we cannot rule out that another implementation of the Gaussian process active learning method could have had better performance, the active sampling framework is less model-dependent and thus is superior for finite population inference. Furthermore, we have proved theoretically that active sampling provides consistent estimators under general conditions. This was also confirmed in our experiments. In our application, we believe that the model-based (Gaussian process) approach to computer experiment emulation is affected by the high complexity of our problem, involving not only one but 44 response surfaces (one per original crash event) that must be learned simultaneously. Yet, this is a fairly small example for scenario generation problems \citep[cf.][]{Ettinger_2021_ICCV, duoba2023tesla} and comprises only a fraction of the crashes in the original crash database \citep{isaksson2005thirty}. Substantial sample sizes would be needed to accurately model all of the response surfaces. Active sampling, targeting a much simpler problem, requires only a rough sketch of the response surface(s) to identify which regions are most informative for estimating the characteristic of interest.

The choice of batch size influenced the performance of the active sampling algorithm, although less so when the number of iterations were large. In practice, one may need to balance the benefits of a smaller batch size on increased statistical efficiency with the benefits of a larger batch size (involving fewer model updates) on increased computational efficiency. With flexible machine learning methods and proper hyper-parameter tuning, carefully avoiding overfitting, we expect this to hold irrespective of the dimension of the problem, although in higher dimensions larger batch sizes may be favored both for computational efficiency and numerical stability. The choice of prediction model had limited influence on performance, as long as the model was flexible enough to capture the true signals in the data. In computer simulation experiment applications, both computational aspects and anticipated performance should be considered for choosing an appropriate model. It is also possible to utilize several machine learning algorithms in the early iterations of the active sampling algorithm to identify the computationally simplest possible model that does not compromise the accuracy of the estimate. Importantly, active sampling was never worse than simple random sampling, even for a misspecified model. Moreover, using an overly complex model (e.g., a non-linear auxiliary model when the true association is linear) only resulted in a minor loss of efficiency of the active sampling estimator. In contrast, ignoring prediction uncertainty resulted in poor performance, particularly in non-linear settings and for misspecified models.

This paper illustrated the active sampling method in an application to generation of simulation scenarios for the assessment of automated emergency braking. In this application, the computation time for running the active sampling algorithm is orders of magnitudes smaller than the computation time for running the corresponding virtual computer experiment simulations. The computational overhead of the training and optimization steps of the active sampling algorithm is thus negligible. The gain in terms of sample size reductions for a given eRMSE therefore translates to a corresponding reduction in total computation time of equal magnitude. The precision obtained by active sampling at $n=2,000$ observations corresponds to an error margin of about $\pm$ 0.5 km/h for the mean impact speed reduction and $\pm 1.0$ percentage points for the crash avoidance rate, which may be considered sufficient in a practical setting. This corresponds to savings of about 95\% in computation time compared to complete enumeration. Not only can the method be applied more broadly in the traffic safety domain, such as for virtual safety assessment of self-driving vehicles of the future, but it can be applied to a wide range of subsampling applications. Future research on the topic may pursue more efficient methods of partitioning the dataset into areas where the outcomes are more precisely predicted or known (where subsampling is less useful) and those where outcomes are less precisely predicted, as well as demonstrate practical applications further.

\section{Conclusion}
\label{sec:conclusion}

We have introduced a machine-learning-assisted active sampling framework for finite population inference, with application to a deterministic computer simulation experiment. We proved theoretically that active sampling provides consistent estimators under general conditions. It was also demonstrated empirically to be robust under different choices of machine learning model. Methods for variance and interval estimation have been proposed, and their validity in the active sampling setting was confirmed empirically. Properly accounting for prediction uncertainty was crucial for the performance of the active sampling algorithm. Substantial performance improvements were observed compared to traditional variance reduction techniques and response surface modeling methods. Active sampling is a promising method for efficient sampling and finite population inference in subsampling applications.

\section*{Acknowledgment}
We would like to thank Volvo Car Corporation for allowing us to use their data and simulation tool, and in particular Malin Svärd and Simon Lundell at Volvo for supporting in the simulation setup. We further want to thank Marina Axelson-Fisk and Johan Jonasson at the Department of Mathematical Sciences, Chalmers University of Technology and University of Gothenburg, for valuable comments on the manuscript.

\section*{Funding}
This research was supported by the European Commission through the SHAPE-IT project under the European Union’s Horizon 2020 research and innovation programme (under the Marie Skłodowska-Curie grant agreement 860410), and in part also by the Swedish funding agency VINNOVA through the FFI project QUADRIS. Also Chalmers Area of Advance Transport funded part of this research.

\section*{Conflict of interest}
The authors report there are no competing interests to declare.

\bigskip
\begin{center}
{\large\bf SUPPLEMENTARY MATERIAL}
\end{center}

\begin{description}

\item[Supplementary Appendix:] Additional theoretical results and proofs (Appendix A), details on the implementation of the sampling methods in the application (Appendix B), and additional experiment results (Appendix C). 

\item[R code and data:] \texttt{R} code and data used for the empirical evaluation in Section \ref{sec:experiments}, application in Section \ref{sec:application}, and replication of main results (Figure 3–5) is available online at \href{https://github.com/imbhe/ActiveSampling}{\nolinkurl{https://github.com/imbhe/ActiveSampling}}.

\end{description}

\bibliographystyle{apalike}
\bibliography{references}

\spacingset{2} 

\appendix
\def\theproposition{S\arabic{proposition}}
\def\thecorollary{S\arabic{corollary}}
\def\theequation{S.\arabic{equation}}
\def\thefigure{S\arabic{figure}}
\def\thetable{S\arabic{table}}
\setcounter{proposition}{0}
\setcounter{corollary}{0}
\setcounter{equation}{0}

\setcounter{figure}{0}

\section{Additional theoretical results and proofs}
\label{appendix:theory}

This appendix contains additional theoretical results and proofs. The optimality of the importance sampling schemes in Proposition \ref{prop:naive_optimality} and \ref{prop:practical_optimality} is proven in Appendix \ref{appendix:optimality}. An asymptotic analysis of the active sampling estimator is presented in Appendix \ref{appendix:clt}.

For a sequence of random variables $\{X, X_n, n \ge 1\}$, we use $X_n \overset{d}{\rightarrow} X$, $X_n \overset{p}{\rightarrow} X$ and $X_n \overset{L_r}{\rightarrow} X$ to denote convergence of $X_n$ to $X$ in distribution, in probability, and in $r^{th}$ mean, respectively.  We will also make use of the following results:
\begin{enumerate}[label = \roman*)]
    \item \textbf{Dominated convergence theorem}: Let $\{X, X_n, n \ge 1\}$ be a sequence of random variables such that $X_n \overset{p}{\rightarrow} X$ and $\E[\sup_{j\ge 1} |X_j|] < \infty$. Then $X_n \overset{L_1}{\rightarrow} X$. 
    \item \textbf{Cramér-Wold theorem}: Let $\boldsymbol{X}, \boldsymbol{X}_1, \boldsymbol{X}_2, \ldots$ be random vectors in $\mathbb{R}^d$. Then $\boldsymbol{X}_n \overset{d}{\rightarrow} \boldsymbol{X}$ if and only if, for every fixed $\boldsymbol{\lambda} \in \mathbb{R}^d$, we have $\boldsymbol{\lambda}^T\boldsymbol{X}_n \overset{d}{\rightarrow} \boldsymbol{\lambda}^T\boldsymbol{X}$.
    \item \textbf{Delta method}: Let $\{\boldsymbol{X}_n\}$ be a sequence of random vectors such that $\sqrt{n}(\boldsymbol{X}_n - \boldsymbol{\tau}_0) \overset{d}{\rightarrow} \mathcal N(\boldsymbol{0}, \boldsymbol{\Gamma}_0)$. Consider a function $h: \mathbb{R}^d \to \mathbb{R}$ and assume that $h(\boldsymbol{u})$ is differentiable in a neighborhood of $\boldsymbol{\tau} = \boldsymbol{\tau}_0$. Then
\begin{equation}
    \label{eq:deltathm}
    \sqrt{n}(h(\boldsymbol{X}_n) - h(\boldsymbol{\tau}_0)) \overset{d}{\rightarrow} \mathcal N\left(0, h(\boldsymbol{\tau}_0)^T\boldsymbol{\Gamma}_0 \nabla h(\boldsymbol{\tau}_0)\right) \quad \text{as } n \rightarrow \infty,
\end{equation}
provided that $h(\boldsymbol{\tau}_0)^T\boldsymbol{\Gamma}_0 \nabla h(\boldsymbol{\tau}_0)>0$. 
\end{enumerate}
For further details, we refer to \citet{Sen1993}.

\clearpage 

\subsection{Optimal importance sampling schemes}
\label{appendix:optimality}

We present in this section proofs of Proposition \ref{prop:naive_optimality} and \ref{prop:practical_optimality}. First, two lemmas are presented.

\begin{lemma}
\label{lemma:multinomial}
\hspace{1cm} \\
Let $\boldsymbol{S} = (S_{1}, \ldots, S_{N}) \sim \mathrm{Multinomial}(n, \boldsymbol{\pi})$, $\boldsymbol{\pi} = (\pi_1, \ldots, \pi_N)$, $\boldsymbol{\mu} := \E[\boldsymbol{S}] = (\mu_{1}, \ldots, \mu_{N})$. Let 
\[
    \hat{\boldsymbol{t}}_{\boldsymbol{y}} = \sum_{i=1}^N S_{i}w_{i}\boldsymbol{y}_i, \quad  w_{i} := \E[S_{i}]^{-1} = (n\pi_{i})^{-1}. 
\]
Then the covariance matrix of $\hat{\boldsymbol{t}}_{\boldsymbol{y}}$ is given by
\[
    \COV(\hat{\boldsymbol{t}}_{\boldsymbol{y}})
    = \frac{1}{n}\left(\sum_{i=1}^N\frac{\boldsymbol{y}_i\boldsymbol{y}_i^T}{\pi_{i}} - \sum_{i,j=1}^N \boldsymbol{y}_i\boldsymbol{y}_j^T \right).
\]
\end{lemma}

\begin{lemma}
\label{lemma:lagrange}
\hspace{1cm} \\
Let $\boldsymbol \pi = (\pi_1, \ldots, \pi_N)$ and consider the function
\begin{equation}
    \label{eq:objective}
    f(\boldsymbol \pi) = \sum_{i = 1}^N\frac{c_i}{\pi_i}  
\end{equation}
for some coefficients $c_i > 0$. Subject to the constraints
\begin{gather*}
    \sum_{i = 1}^N \pi_i = 1 , \quad \pi_i > 0 ,
\end{gather*}
$f(\boldsymbol \pi)$ is minimized by $\boldsymbol{\pi}^* = (\pi_1^*, \ldots, \pi_N^*)$ with
\[
    \pi_i^* = \frac{\sqrt{c_i}}{\sum_{j = 1}^N \sqrt{c_j}} , \quad i = 1, \ldots, N.
\]
\end{lemma}

\begin{proof}{of Lemma \ref{lemma:multinomial}}
\hspace{1cm} \\
By properties of the multinomial distribution, we have that
\[
    \mu_{i} := \E[S_{i}] = n\pi_{i}, \quad
    \Var(S_{i}) = n\pi_{i}(1-\pi_{i}), \quad \Cov(S_{i}, S_{j}) = - n\pi_{i}\pi_{j}, 
\]
for $i,j = 1, \ldots, N, i \ne j$. Hence, 
\begin{align*}
    \COV(\hat{\boldsymbol{t}}_{\boldsymbol{y}}) 
    & = \COV\left(\sum_{i=1}^N S_iw_i\boldsymbol{y}_i\right) 
    =\left(\sum_{i=1}^N \frac{n\pi_{i}(1-\pi_{i})}{n^2\pi_{i}^2} \boldsymbol{y}_i\boldsymbol{y}_i^T - \sum_{\overset{i,j=1}{i \ne j}}^N \frac{n\pi_{i}\pi_{j}}{n^2\pi_{i}\pi_{j}} \boldsymbol{y}_i\boldsymbol{y}_j^T \right) \\
    & = \frac{1}{n}\left(\sum_{i=1}^N\frac{1-\pi_{i}}{\pi_{i}} \boldsymbol{y}_i\boldsymbol{y}_i^T - \sum_{\overset{i,j=1}{i \ne j}}^N \boldsymbol{y}_i\boldsymbol{y}_j^T \right) 
     = \frac{1}{n}\left(\sum_{i=1}^N\frac{\boldsymbol{y}_i\boldsymbol{y}_i^T}{\pi_{i}} - \sum_{i,j=1}^N \boldsymbol{y}_i\boldsymbol{y}_j^T \right).
\end{align*}
\hfill $\square$
\end{proof}

\begin{proof}{of Lemma \ref{lemma:lagrange}}
\hspace{1cm} \\
Using the method of Lagrange multipliers \citep{Boyd2004}, we introduce the auxiliary function
\begin{gather*}
    \Lambda(\boldsymbol \pi, \lambda) = f(\boldsymbol \pi)+\lambda g(\boldsymbol \pi) , \quad  g(\boldsymbol \pi) = \sum_{i =1}^N \pi_i-1\text{ }.
\end{gather*}
Critical points of the Lagrangian are found by solving the equation system
\begin{align*}
    \nabla \Lambda(\boldsymbol  \pi, \lambda) = \mathbf{0} \quad
    \Leftrightarrow \quad
    \begin{cases} 
    g(\boldsymbol \pi) = 0 \\ 
    -\nabla_{\boldsymbol \pi} f(\boldsymbol \pi) = \lambda \nabla_{\boldsymbol \pi} g(\boldsymbol \pi) 
    \end{cases}\text{ }.
\end{align*}
Since $\frac{\partial f(\boldsymbol \pi)}{\partial \pi_i} = -c_i/\pi_i^2$ and $\frac{\partial g(\boldsymbol \pi)}{\partial \pi_i} = 1$, this implies that $\lambda = c_1/\pi_1^2 = \ldots = c_N/\pi_N^2$, and further that 
$
|\pi_i| \propto |\sqrt{c_i}|.
$
Since $c_i > 0$, $\pi_i > 0$ and $\sum_{i = 1}^N \pi_i = 1$, we obtain 
\begin{equation}
    \label{eq:pik}
    \pi_i^* =\frac{\sqrt{c_i}}{\sum_{j = 1}^N \sqrt{c_j}}, \quad {i = 1, \ldots, N}.
\end{equation}
Thus, the point  $(\boldsymbol \pi^*, \lambda^*)$ with entries $\pi_i^*$ defined according to \eqref{eq:pik} and $\lambda^* = c_1 / \pi_1^{*2}$ is a stationary point of $\Lambda(\boldsymbol \pi, \lambda)$. Hence, $\boldsymbol \pi^*$ is a stationary point of $f(\boldsymbol \pi)$ under the specified constraints. Specifically, $\boldsymbol \pi^*$ is a local minimum. Since we consider a convex function over a convex set, $\boldsymbol \pi^*$ also is a global minimum. This proves the optimality of \eqref{eq:pik}. \hfill $\square$
\end{proof}

\begin{proof}{of Proposition \ref{prop:naive_optimality}}
\hspace{1cm} \\
Let $\hat{\boldsymbol{t}}_{\boldsymbol{y},k} = \sum_{i=1}^N S_{ki} w_{ki}\boldsymbol{y}_i$, $w_{ki} = (m_k\pi_{i})^{-1}$, and $\hat{\theta}_k = h(\hat{\boldsymbol{t}}_{\boldsymbol{y},k})$. Since $\boldsymbol{S}_{k}\sim\mathrm{Multinomial}(m_k, \boldsymbol{\pi})$, it can be written as the sum $\boldsymbol{S}_k = \boldsymbol{X}_1+\ldots+\boldsymbol{X}_{m_k}$ of $m_k$ independent multinoulli random variables $\boldsymbol{X}_j \sim \mathrm{Multinomial}(1, \boldsymbol{\pi})$, $j=1, \ldots, k$. The estimator $\hat{\boldsymbol{t}}_{\boldsymbol{y},k}$ can therefore also be written as the sum $\hat{\boldsymbol{t}}_{\boldsymbol{y},k} = \sum_{j=1}^{m_k} \boldsymbol{T}_j$ of $m_k$ independent random variables $\boldsymbol{T}_j = \frac{1}{m_k}\sum_{i=1}^N \pi_i^{-1}X_{ji} \boldsymbol{y}_i$ with common mean $m_k^{-1}\boldsymbol{t}_{\boldsymbol{y}}$. Since $\{\boldsymbol{y}_i\}_{i=1}^N$ and $\boldsymbol{\pi}$ are fixed, the covariance matrices of $\{\boldsymbol{T}_j\}_{j\ge 1}$ are all finite. By the central limit theorem for triangular arrays of independent random variables, it follows that
\[
\sqrt{m_k} (\hat{\boldsymbol{t}}_{\boldsymbol{y},k} - \boldsymbol{t}_{\boldsymbol{y}}) \overset{d}{\rightarrow} \mathcal N(\boldsymbol{0}, \boldsymbol{\Gamma}_0) \quad \text{ as } k \rightarrow \infty.
\]
The asymptotic covariance matrix $\boldsymbol{\Gamma}_0$ is obtained from Lemma \ref{lemma:multinomial} as
\[
\boldsymbol{\Gamma}_0 = \left(\sum_{i=1}^N\frac{\boldsymbol{y}_i\boldsymbol{y}_i^T}{\pi_{i}} - \sum_{i,j=1}^N \boldsymbol{y}_i\boldsymbol{y}_j^T \right).
\]
By the delta method \eqref{eq:deltathm}, it follows that
\[
\sqrt{m_k}(\hat{\theta}_k - \theta) \overset{d}{\rightarrow} \mathcal N(0, \gamma^2) \quad \text{as } k \rightarrow \infty,
\]
with $\gamma^2 = \nabla h(\boldsymbol{t}_{\boldsymbol{y}})^T\boldsymbol{\Gamma}_0\nabla h(\boldsymbol{t}_{\boldsymbol{y}})$. The asymptotic mean squared error of $\hat{\theta}$ is thus given by
\begin{equation}
    \label{eq:amse}
    \mathrm{AMSE}(\hat\theta) = 
    \nabla h(\boldsymbol{t}_{\boldsymbol{y}})^T
    \left(\sum_{i=1}^N\frac{\boldsymbol{y}_i\boldsymbol{y}_i^T}{\pi_{i}} - \sum_{i,j=1}^N \boldsymbol{y}_i\boldsymbol{y}_j^T \right)
    \nabla h(\boldsymbol{t}_{\boldsymbol{y}}).
\end{equation}
To minimize \eqref{eq:amse}, it suffices to minimize
\[
\nabla h(\boldsymbol{t}_{\boldsymbol{y}})^T\sum_{i=1}^N\frac{\boldsymbol{y}_i\boldsymbol{y}_i^T}{\pi_{i}}\nabla h(\boldsymbol{t}_{\boldsymbol{y}}),
\]
which can be written on the form \eqref{eq:objective} with $c_i = \bigr\rvert\nabla h(\boldsymbol{u})^T\boldsymbol{y}_i\bigr\rvert^2_{\boldsymbol{u} = \boldsymbol{t}_{\boldsymbol{y}}}$. The desired result now follows from Lemma \ref{lemma:lagrange}. \hfill $\square$ 
\end{proof}

\begin{proof}{of Proposition \ref{prop:practical_optimality}}
\hspace{1cm} \\
The asymptotic mean squared error $\mathrm{AMSE}(\hat\theta)$ was given in the proof of Proposition \ref{prop:naive_optimality} by \eqref{eq:amse}. Substituting the unknown constants $\boldsymbol{y}_i$ with the random variables $\boldsymbol{Y}_i$, we obtain the expected asymptotic mean squared error of $\hat{\theta}$ as 
\begin{equation}
    \label{eq:eamse}
    \E_{\boldsymbol{Y}}[\mathrm{AMSE}(\hat\theta)] = 
    \E\left[
    \nabla h(\boldsymbol{u})^T
    \left(\sum_{i=1}^N\frac{\boldsymbol{Y}_i\boldsymbol{Y}_i^T}{\pi_{i}} - \sum_{i,j=1}^N \boldsymbol{Y}_i\boldsymbol{Y}_j^T \right)
    \nabla h(\boldsymbol{u}).
    \right]_{\boldsymbol{u} = \boldsymbol{t}_{\boldsymbol{y}}}.
\end{equation}
To minimize \eqref{eq:eamse}, it suffices to minimize
\begin{equation}
    \label{eq:objective_eamse}
    \sum_{i=1}^N\frac{\E\left[
    h(\boldsymbol{u})^T
    \boldsymbol{Y}_i\boldsymbol{Y}_i^T
    \nabla h(\boldsymbol{u})
    \right]_{\boldsymbol{u} = \boldsymbol{t}_{\boldsymbol{y}}}}{\pi_{i}}. 
\end{equation}
Using the equality $\E[X^2] = \E[X]^2 + \Var(X)$, we have that
\begin{align}
    \nonumber
    \E\left[(\nabla h(\boldsymbol{u})^T \boldsymbol{Y}_i)^2\right] 
    & = \E[\nabla h(\boldsymbol{u})^T \boldsymbol{Y}_i]^2 
    + \Var\left(\nabla h(\boldsymbol{u})^T \boldsymbol{Y}_i\right) \\
    \nonumber
    & = (\nabla h(\boldsymbol{u})^T \E[\boldsymbol{Y}_i])^2 
    + \nabla h(\boldsymbol{u})^T\COV\left( \boldsymbol{Y}_i\right)\nabla h(\boldsymbol{u}) \\
    \label{eq:simplification}
    & = (\nabla h(\boldsymbol{u})^T \boldsymbol{\eta}_i)^2 + \nabla h(\boldsymbol{u})^T\boldsymbol{\Sigma}_i \nabla h(\boldsymbol{u}).
\end{align}
Inserting \eqref{eq:simplification} into \eqref{eq:objective_eamse}, the result now follows Lemma \ref{lemma:lagrange} with \newline $c_i = \left[(\nabla h(\boldsymbol{u})^T\boldsymbol{\eta}_i)^2 + \nabla h(\boldsymbol{u})^T \boldsymbol{\Sigma}_i \nabla h(\boldsymbol{u})\right]_{\boldsymbol{u} = \boldsymbol{t}_{\boldsymbol{y}}}$. \hfill $\square$
\end{proof}

\clearpage

\subsection{Central limit theorems} 
\label{appendix:clt}

We provide in Proposition \ref{prop:martingaleCLT} conditions under which the active sampling estimator $\hat t_y^{(k)}$ of a scalar total $t_y$ is consistent and asymptotically normally distributed, and present consistent variance estimators. A generalization to multivariate estimators and to characteristics defined as smooth functions of totals is provided in Corollary \ref{cor:multivariateMartingaleCLT}. To show asymptotic normality of our active sampling estimator, we use the following result of \citet{Brown1971}: 

\begin{lemma}[Martingale central limit theorem]
\label{lemma:martingaleCLT}
\hspace{0.1cm} \\
Consider a sequence $\{X_j\}_{j=1}^{\infty}$ of random variables such that $\E[X_j] = \E[X_j|X_1, \ldots, X_{j-1}] = 0$ and $\E[X_j^2] < \infty$. Let $\sigma_j^2 = \E[X_j^2|X_1, \ldots, X_{j-1}]$, $U_k = \sum_{j=1}^k X_j$, $V_k^2 = \sum_{j=1}^k \sigma_j^2$, and $u_k^2 = \E[U_k^2] = \E[V_k^2]$. Assume that $V_k^2u_k^{-2} \overset{p}{\rightarrow} 1$ as $k \rightarrow \infty$,
and that the Lindeberg-Feller condition holds:
\begin{equation}
    \label{eq:Lindeberg-Feller}
    u_k^{-2}\sum_{j=1}^k \E[X_j^2 I(|X_j|>\varepsilon u_k)] \rightarrow 0 \quad \text{ as } k \rightarrow \infty \quad \text{ for all } \varepsilon > 0.
\end{equation} 
Then
\[
    U_k / u_k \overset{d}{\rightarrow} \mathcal{N}(0,1) \quad \text{as } k \rightarrow \infty.
\]
\end{lemma}
A corresponding central limit theorem for active sampling is presented below. 

\begin{proposition}[Central limit theorem, $n_k$ bounded, $k \rightarrow \infty$]
\label{prop:martingaleCLT}
\hspace{1cm} \\
Consider a finite index set $\mathcal D = \{1, \ldots, N\}$ with corresponding data $y_1, \ldots, y_N$, and infinite sequence $\{n_j\}_{j=1}^{\infty}$ with $n_j \in \mathbb{N}$, $n_j < N$. Let $\{\boldsymbol{S}_{j}\}_{j=1}^{\infty}$ be an infinite sequence of random vectors $\boldsymbol{S}_{j} = (S_{j1}, \ldots, S_{jN}) \in \mathbb{N}^N$ such that $\sum_{i=1}^{N} \mu_{ji} = n_j$, where $\mu_{ji} := \E[S_{ji}|\boldsymbol{S}_1, \ldots \boldsymbol{S}_{j-1}]$ are assumed to be strictly positive for all $j,i$. Let $t_y = \sum_{i\in \mathcal D}y_i$, $\hat t_{y,j} = \sum_{i\in \mathcal D}\frac{S_{ji}y_i}{\mu_{ji}}$, $m_k = \sum_{j=1}^k n_j$, $\hat t_{y}^{(k)} = \frac{1}{m_k}\sum_{j = 1}^k n_j\hat t_{y,j}$, $\sigma_j^2 = \Var(\hat t_{y,j}|\boldsymbol{S}_1, \ldots, \boldsymbol{S}_{j-1})$, $A_k^2 = \sum_{j=1}^k n_j^2 \sigma_j^2$, and $b_k^2 = \Var(\sum_{j=1}^k n_j\hat t_{y,j})$. Assume that
\begin{enumerate}[label=(A\arabic*)]
    \item $\quad S_{ji}/\mu_{ji}$ have uniformly bounded second moments,
    \item $\quad b_k \rightarrow \infty$ as $k \rightarrow \infty$, and
    \item $\quad A_k^2b_k^{-2} \overset{p}\rightarrow 1$ as $k \rightarrow \infty$.
\end{enumerate}
Then
\begin{align}
\label{eq:martingaleCLT}
    \frac{\hat t_y^{(k)} - t_y}{b_k/m_k} \overset{d}{\rightarrow} \mathcal N(0, 1) \quad & \text{ as } k \rightarrow \infty, \quad \text{and} \\ 
    \label{eq:martingaleVarEst}
    b_k^{-2}\sum_{j=1}^k n_j^2\left(\hat t_{y,j}-\hat t_y^{(k)}\right)^2 \overset{p}{\rightarrow} 1 \quad & \text{ as } k \rightarrow \infty. 
\end{align}
Furthermore, if $\hat{\sigma}_{j}^2$ are unbiased estimators of the conditional variances $\sigma_j^2$, i.e.,\\ $\E[\hat{\sigma}_{j}^2|\boldsymbol{S}_1, \ldots, \boldsymbol{S}_{j-1}] = \sigma_j^2$, and 
\begin{enumerate}[label=(A\arabic*)]
    \setcounter{enumi}{3}
    \item $\quad b_k^{-2}\Var(\sum_{j=1}^k n_j^2 \hat{\sigma}_j^2)$ are uniformly bounded, 
\end{enumerate}
then additionally we have that
\begin{align}
    \label{eq:classical_var_est}
    b_k^{-2}\sum_{j=1}^k n_j^2 \hat{\sigma}_j^2 \overset{p}{\rightarrow} 1 \quad & \text{ as } k \rightarrow \infty.    
\end{align}
\end{proposition}

The first result \eqref{eq:martingaleCLT} establishes asymptotic normality of the active learning estimator $\hat t_y^{(k)}$ under the specified conditions. We note that $b_k = O(m_k^{1/2})$, so the convergence of $\hat t_y^{(k)}$ to $t_y$ is at the usual parametric rate $m_k^{-1/2}$. The second result \eqref{eq:martingaleVarEst} proves the consistency of the martingale variance estimator $m_k^{-1}\sum_{j=1}^k n_j^2\left(\hat t_{y,j}-\hat t_y^{(k)}\right)^2$, and the third \eqref{eq:classical_var_est} consistency of the design-based variance estimator $m_k^{-1}\sum_{j=1}^k n_j^2 \hat{\sigma}_j^2$ (cf. Section \ref{sec:variance_estimation}).

Since, in any sensible probability sampling design, $S_{ji}$ have finite second moments, the first assumption (A1) is fulfilled if the sampling probabilities (and corresponding means $\mu_{ji}$) are properly bounded away from zero. The second assumption (A2) requires the total variance $\Var(\sum_{j=1}^k n_j\hat t_{y,j})$ to tend to infinity with $k$. This may at first sight seem to contradict the purpose of active sampling, which is to make the variance as small as possible. Indeed, if all the $y_i$'s have the same sign it is theoretically possible to construct a sampling strategy that produces an estimator with zero variance, which clearly does not converge to a normal limit. In practice, however, finding the true optimal design is not possible and a sampling strategy with good performance generally also fulfills the assumptions (A1) and (A2).

The third assumption states that the sum of conditional variances asymptotically should behave like the total variance. Hence, the statistical properties of the active sampling estimator can be deduced from a single execution of the algorithm. Empirical justification for this assumption is provided in Section \ref{sec:application}. We note that the fourth assumption (A4), needed for consistency of the classical variance estimator, is stronger than the second (A2). To see this, note that (A4) requires $\hat\sigma_{j}^2$ to have bounded second moments for every $j$. But $\hat\sigma_{j}^2$ depends on $S_{ji}/\mu_{ji}^2$, which is larger than $S_{ji}/\mu_{ji}$ for all $j,i$ such that $\mu_{ji}\le 1$, as is the case for all or nearly all $j,i$ in all realistic subsampling applications. For fixed-size designs $\hat\sigma_{j}^2$ also depend on the joint selection probabilities, which means that $\E[S_{ji}S_{jl}]$ need to be properly bounded away from zero for consistent variance estimation. Note, in particular, that this requires all $n_j \ge 2$ for fixed-size designs, whereas (A2) makes no such restriction.

Before providing a proof, we present below a generalization to vectors of totals and smooth functions of totals.

\begin{corollary}[Multivariate central limit theorem, $n_k$ bounded, $k \rightarrow \infty$]
\label{cor:multivariateMartingaleCLT}
\hspace{1cm} \\
Consider a finite index set $\mathcal D = \{1, \ldots, N\}$ with corresponding data $\boldsymbol{y}_1, \ldots, \boldsymbol{y}_N \in \mathbb{R}^d$. Let $\{n_j\}_{j=1}^{\infty}$, $\{\boldsymbol{S}_{j}\}_{j=1}^{\infty}$, $\mu_{ji}$ and $m_k$ be defined as in Proposition \ref{prop:martingaleCLT}. Let $\boldsymbol{t}_{\boldsymbol{y}} = \sum_{i\in \mathcal D} \boldsymbol{y}_i$, $\hat{\boldsymbol{t}}_{\boldsymbol{y},j} = \sum_{i\in \mathcal D}\frac{S_{ji}\boldsymbol{y}_i}{\mu_{ji}}$, and $\hat{\boldsymbol{t}}_{\boldsymbol{y}}^{(k)} = \frac{1}{m_k}\sum_{j = 1}^k n_j\hat{\boldsymbol{t}}_{\boldsymbol{y},j}$. Consider a function $h: \mathbb{R}^d \to \mathbb{R}$, and assume that $h(\boldsymbol{u})$ is differentiable in a neighborhood of $\boldsymbol{u} = \boldsymbol{t}_{\boldsymbol{y}}$, with $\nabla h(\boldsymbol{u})\rvert_{\boldsymbol{u} = \boldsymbol{t}_{\boldsymbol{y}}} \ne \boldsymbol{0}$. Let $\boldsymbol{\Phi}_{j} = \COV(\hat{\boldsymbol{t}}_{y,j}|\boldsymbol{S}_1, \ldots, \boldsymbol{S}_{j-1})$, $\boldsymbol{A}_{k} = \sum_{j=1}^k n_j^2 \boldsymbol{\Phi}_{j}$ and $\boldsymbol{B}_{k} = \COV(\sum_{j=1}^k n_j\hat{\boldsymbol{t}}_{y,j})$. Assume that 
\begin{enumerate}[label=\roman*)]
    \item $S_{ji}/\mu_{ji}$ have uniformly bounded second moments,
    \item $m_k^{-1}\boldsymbol{B}_{k}$ converges elementwise to some matrix $\boldsymbol{\Psi}_0$, and $\boldsymbol{\Psi}_0$ is full rank,
    \item $\boldsymbol{\lambda}^T\boldsymbol{B}_{k}\boldsymbol{\lambda} \rightarrow \infty$ as $k \rightarrow \infty$ for every $\boldsymbol{\lambda} \in \mathbb{R}^d \setminus \boldsymbol{0}$, and
    \item $\boldsymbol{A}_{k} \boldsymbol{B}_{k}^{-1} \overset{p}{\rightarrow} \boldsymbol{I}_{d\times d}$ (elementwise) as $k \rightarrow \infty$.
\end{enumerate}
Then
\begin{align*}
    \sqrt{m_k}\left(\hat{\boldsymbol{t}}_{\boldsymbol{y}}^{(k)} - \boldsymbol{t}_{\boldsymbol{y}}\right) \overset{d}{\rightarrow} \mathcal N(\boldsymbol{0}, \boldsymbol{\Psi}_0) \quad & \text{ as } k \rightarrow \infty, \quad \text{and} \\
    \sqrt{m_k}\left(h(\hat{\boldsymbol{t}}_{\boldsymbol{y}}^{(k)}) - h(\boldsymbol{t}_{\boldsymbol{y}})\right) \overset{d}{\rightarrow} \mathcal N(\boldsymbol{0}, \gamma_0^2) \quad & \text{ as } k \rightarrow \infty,
\end{align*}
provided that $\gamma_0^2>0$, where $\gamma_0^2 = \nabla h(\boldsymbol{u})^T\boldsymbol{\Psi}_0\nabla h(\boldsymbol{u})\rvert_{\boldsymbol{u} = \boldsymbol{t}_{\boldsymbol{y}}}$. Moreover, the asymptotic covariance matrix $\boldsymbol{\Psi}_0$ and variance $\gamma_0^2$ can be consistently estimated by 
\begin{align*}
    \hat{\boldsymbol{\Psi}}^{(k)} & = \frac{1}{m_k^2}\sum_{j=1}^k n_j^2\left(\hat{\boldsymbol{t}}_{\boldsymbol{y},j}-\hat{\boldsymbol{t}}_{\boldsymbol{y}}^{(k)}\right)\left(\hat{\boldsymbol{t}}_{\boldsymbol{y},j}-\hat{\boldsymbol{t}}_{\boldsymbol{y}}^{(k)}\right)^T, \\
    \hat{\gamma}^2_k & = \nabla h(\boldsymbol{u})^T\hat{\boldsymbol{\Psi}}^{(k)}\nabla h(\boldsymbol{u})\rvert_{\boldsymbol{u} = \hat{\boldsymbol{t}}_{\boldsymbol{y}}^{(k)}}.
\end{align*}
Furthermore, if $\hat{\boldsymbol{\Phi}}_{j}$ are unbiased estimators of the conditional covariance matrices $\boldsymbol{\Phi}_j$, i.e., $\E[\hat{\boldsymbol{\Phi}}_{j}|\boldsymbol{S}_1, \ldots, \boldsymbol{S}_{j-1}] = \boldsymbol{\Phi}_j$, and
\begin{enumerate}[label=\roman*)]
    \setcounter{enumi}{3}
    \item $\quad (\boldsymbol{\lambda}^T\boldsymbol{B}_k\boldsymbol{\lambda})^{-1} \Var(\sum_{j=1}^k\boldsymbol{\lambda}^T\hat{\boldsymbol{\Phi}}_{j}\boldsymbol{\lambda})$ are uniformly bounded for every $\boldsymbol{\lambda} \in \mathbb{R}^d \setminus \boldsymbol{0}$, 
\end{enumerate}
then the asymptotic covariance matrix $\boldsymbol{\Psi}_0$ and variance $\gamma_0^2$ can also be consistently estimated by 
\begin{align*}
    \widetilde{\boldsymbol{\Psi}}^{(k)} & = \frac{1}{m_k^2}\sum_{j=1}^k n_j^2 \hat{\boldsymbol{\Phi}}_{j},  \\
    \widetilde{\gamma}^{2}_k & = \nabla h(\boldsymbol{u})^T\widetilde{\boldsymbol{\Psi}}^{(k)} \nabla h(\boldsymbol{u})\rvert_{\boldsymbol{u} = \hat{\boldsymbol{t}}_{\boldsymbol{y}}^{(k)}}.
\end{align*}
\end{corollary}

\begin{proof}{of Proposition \ref{prop:martingaleCLT}}
\hspace{1cm} \\
Let $X_j = n_j(\hat t_{y,j} - t_y)$, $U_k = \sum_{j=1}^k X_j$,
$V_k^2 = \sum_{j=1}^k \E[X_j^2 | X_1, \ldots, X_{j-1}] = A_k^2$, and $u_k^2 = \E[U_k^2] = \E[V_k^2] = b_k^2$. Note that $\E[X_j] = 0$, and that $X_j$ by (A1) have uniformly bounded second moments, and hence that $\max_{j \le k} \E[X_j^2]$ are uniformly bounded. Since $u_k \rightarrow \infty$ as $k \rightarrow \infty$, we therefore have that $\max_{j \le k} u_k^{-1}X_j \overset{L_2}{\rightarrow} 0$, which implies 
\begin{equation}
    \label{eq:maxto0}
    \max_{j \le k} u_k^{-1}X_j \overset{p}{\rightarrow} 0.
\end{equation}
This in turn is equivalent to the weaker Lindeberg-Feller condition 
\begin{equation}
    \label{eq:weakL-F}
    u_k^{-2}\sum_{j=1}^k X_j^2 I(|X_j|\ge \varepsilon u_k) \overset{p}{\rightarrow} 0 \quad \text{ for all } \varepsilon > 0,   
\end{equation}
since $P(\max_{j \le k} u_k^{-1}X_j > \varepsilon) = P(\sum_{j=1}^k u_k^{-2} X_j^2 I(|X_j| > \varepsilon u_k) > \varepsilon^2)$. But 
\begin{equation}
    \label{eq:dom}
    \E\left[\bigr\rvert u_k^{-2}\sum_{j=1}^k X_j^2 I(|X_j|\ge \varepsilon u_k)\bigr\rvert\right] \le u_k^{-2} \E\left[\sum_{j=1}^k X_j^2\right] = 1 \quad \text{for all } k.
\end{equation}
By the dominated convergence theorem, \eqref{eq:weakL-F} and \eqref{eq:dom} implies the Lindeberg-Feller condition \eqref{eq:Lindeberg-Feller}, which together with (A3) according to Lemma \ref{lemma:martingaleCLT} gives
\[
    U_k / u_k  \overset{d}{\rightarrow} \mathcal N(0, 1) \quad \text{ as } k \rightarrow \infty.
\]
The first result \eqref{eq:martingaleCLT} now follows by noting that
\[
    U_k / u_k  = \frac{\sum_{j=1}^k n_j(\hat t_{y,j} - t_y)}{b_k} = \frac{m_k^{-1}\sum_{j=1}^k n_j(\hat t_{y,j} - t_y)}{b_k / m_k} = \frac{\hat t_y^{(k)} - t_y}{b_k/m_k}.
\]
For \eqref{eq:martingaleVarEst}, we first note that $\hat t_{y,k}= O_p(1)$ and $\hat t_y^{(k)} = t_y + O_p(b_k/m_k)$. Hence 
\begin{align*}
    b_k^{-2}\sum_{j=1}^k n_j^2\left(\hat t_{y,j}-\hat t_y^{(k)}\right)^2 
    & = b_k^{-2}\sum_{j=1}^k n_j^2 \left(\hat t_{y,j}-t_y + O_p(b_k/m_k)\right)^2 \\
    & = b_k^{-2}\sum_{j=1}^kn_j^2\left(\hat t_{y,j}-t_y\right)^2 + O_p(b_k^{-1}).  
\end{align*}
Next, 
\begin{align*}
    \E\biggr\rvert b_k^{-2}\sum_{j=1}^k n_j^2\left(\hat t_{y,j}-\hat t_y^{(k)}\right)^2  - 1\biggr\rvert
    & = \E\biggr\rvert b_k^{-2}\sum_{j=1}^k n_j^2\left(\hat t_{y,j}-t_y \right)^2 + O_p(b_k^{-1}) - 1 \biggr\rvert \\
    & \le \E\biggr\rvert b_k^{-2}\sum_{j=1}^kn_j^2\left(\hat t_{y,j}-t_y \right)^2 - 1\biggr\rvert + \E[O_p(b_k^{-1})].
\end{align*}
Note now that (A3) is equivalent to $\lim_{k\rightarrow \infty}\E|A_k^2b_k^{-2} -1|$ \citep[][Lemma 1]{Brown1971}, which together with \eqref{eq:maxto0} and the Lindeberg-Feller condition \eqref{eq:Lindeberg-Feller} implies that the first term vanishes as $k \rightarrow \infty$ \citep[][Theorem 3.5]{Hall1980}. As does the second term, since $\hat t_{y,j}$ have uniformly bounded second moments. Hence, \eqref{eq:martingaleVarEst} now follows since convergence in mean implies convergence in probability.

For the last result \eqref{eq:classical_var_est}, we note that $\E[\sum_{j=1}^k n_j^2 \hat{\sigma}_j^2] = b_k^2$, and 
\[
    \Var\left(b_k^{-2}\sum_{j=1}^k n_j^2 \hat{\sigma}_j^2\right) = 
    \frac{\Var(\sum_{j=1}^k n_j^2 \hat{\sigma}_j^2)}{b_k^2}\frac{1}{b_k^2} \rightarrow 0 \quad \text{as } k \rightarrow \infty,
\]
since the first factor by (A4) is bounded and the second goes to zero as $k \rightarrow \infty$. This completes the proof.
\hfill $\square$
\end{proof}

\begin{proof}{of Corollary \ref{prop:martingaleCLT}}
\hspace{1cm} \\
The result follows immediately from Proposition \ref{prop:martingaleCLT} by application of the Cramér-Wold theorem and the Delta method \citep{Sen1993} . 
\end{proof}

\clearpage
\section{Additional method and implementation details}
\label{appendix:methods}

Additional details on the implementation of the sampling methods in the application (Section \ref{sec:application}) are provided below.

\subsection{Active sampling} 

\paragraph{Optimal sampling schemes} Optimal sampling schemes for active sampling in the crash-causation-based scenario generation application are derived below. Note first that the characteristic of interest (7) can be written on the form (1) with $\boldsymbol{y}_i = p_i(r_i, r_i x_i)^T$ and $h(\boldsymbol{u}) = u_2 / u_1$, where $r_i$ is a binary indicator taking the value $1$ if there was a crash in the baseline scenario and $0$ otherwise, $p_i$ the prior observation weight (scenario probability), and $x_i$ the impact speed reduction or crash avoidance indicator with the AEB compared to baseline driving (without the AEB). The observation weights $p_i$ are known \textit{a priori} and need not be learned from data, whereas $r_i$ and $x_i$ can only be observed by running the corresponding virtual simulation. We therefore introduce the random variables $R_i, X_i$ to describe our uncertainty in the variables of interest ($r_i, x_i$), and $\boldsymbol{Y}_i = p_i(R_i, R_i X_i)^T$. Since the two components of $\boldsymbol{Y}_i$ are correlated through the common factor $R_i$ and $\boldsymbol{Y}_i = \boldsymbol{0}$ if $R_i = 0$, we model $(X_i, R_i)$ as 
\begin{align*}
    R_i | \boldsymbol{z}_i & \sim f(r_i|\boldsymbol{z}_i) \\
    X_i | R_i, \boldsymbol{z}_i & \sim f(x_i|r_i, \boldsymbol{z}_i).
\end{align*}
Using the law of total expectation and total covariance (conditioning on $R_i$) to evaluate the expectation and covariance matrix of $\boldsymbol{Y}_i$, we obtain the optimal sampling scheme for the active sampling algorithm as
\begin{equation}
    \label{eq:is_application}
    \pi_{ki} \propto \sqrt{c_i}, \quad c_i = p_i^2\hat r_i \left[(\hat x_i - \hat\theta^{(k-1)})^2 + \hat\sigma_i^2\right],  
\end{equation}
where $\hat r_i$ is the predicted probability of generating a crash in the baseline scenario, $\hat x_i$ the predicted impact speed reduction or probability of crash avoidance with the AEB given that there was a crash in the baseline scenario, $\sigma_i^2$ the residual error (root mean squared prediction error for regression, standard deviation $\sqrt{\hat x_i (1 - \hat x_i)}$ for classification), and $\hat\theta^{(k-1)}$ the estimate of the characteristic of interest from the previous iteration of the algorithm. In the light of \eqref{eq:is_application}, it is optimal in the absence of prior information about $r_i$ and $x_i$ to set $\pi_{ki}\propto p_i$. Hence, this was also used in the initial iteration(s) of the active sampling algorithm until reliable predictions were obtained, i.e., until the prediction R-squared (regression) or prediction accuracy (classification) was greater than zero when evaluated on hold-out data.

\paragraph{Learning and prediction} For the learning step of the active sampling algorithm, we used random forest regression for continuous outcomes (impact speed reduction) and random forest classification for binary outcomes (crash/no-crash under baseline and countermeasure scenarios) \citep{Breiman2001}. The random forest method was chosen for the following reasons: i) it is fast and flexible, ii) it is capable of finding non-linear and non-monotonic patterns, as well as interactions between variables, without the need for explicit feature construction, and iii) measures of generalization error and prediction performance are readily available through estimates of residual variance, prediction R-squared and accuracy on hold-out (out-of-bag) data.

Explanatory variables were the input parameters to the simulations (off-road glance duration and maximal deceleration), and an \textit{a priori} known case-specific maximal impact speed. The maximal impact speed could be retrieved by running a single simulation per original crash event at the maximal glance duration and minimum deceleration. Although in reality this would count as 44 extra simulations (one per original crash event) it has only a minor effect on the overall performance evaluation and hence not accounted for in the presentation of the results.

Random forests were fitted using 100 trees with variance splitting rule for regression, and gini splitting rule for classification. Other hyper-parameters (minimum node size and number of variables to split upon) were selected with 5-fold cross validation, using a random grid search of minimum node size from 1 to 20 and number of variables to split upon from 1 to 3. All predictions were set equal if the model could not be fitted or produced a prediction R-squared less than 0 on hold-out data, thus resorting to density importance sampling ($\pi_{ki} \propto p_i$). To reduce computation time, prediction models were updated every $10^{\mathrm{th}}$ new observation up to a sample size of $n=100$ observations, thereafter every $25^{\mathrm{th}}$ new observation up to a sample size of $n=500$ observations, thereafter every $50^{\mathrm{th}}$ observation up to a sample size of $n=1,000$ observations, and so on.

We also performed sensitivity analyses for the choice of machine learning algorithm using extreme gradient boosting \citep[XGBoost,][]{Chen2016} and k-nearest neighbors (k-NN). As before, the hyper-parameters (learning rate, maximum tree-depth and number of boosting rounds for XGBoost, and number of neighbors for k-NN) were tuned by using cross-validation.

\subsection{Importance sampling}

Two different importance sampling schemes were considered: density importance sampling and severity importance sampling. With density importance sampling, the sampling probabilities were selected proportional to the prior observation weights $p_i$, since instances with large observation weights by design of the scenario generation framework have a larger contribution to estimation. Since our aim was safety benefit evaluation of an advanced driver assistance system compared to a baseline driving scenario, and the potential safety benefit increases with impact severity, we hypothesized that oversampling of high-severity instances would lead to further variance reduction in the safety benefit evaluation. We therefore also included severity importance sampling, which attempts to oversample high-severity instances by assigning sampling probabilities proportional to $p_i \times o_i \times  d_i \times m_i$, where $p_i$ is the prior observation weight of instance $i$, $o_i$ is the corresponding off-road glance duration, $d_i$ is the maximal deceleration, and $m_i$ an \textit{a priori} known maximal possible impact speed of instance $i$. To account for scale differences between variables, all variables (off-road glance duration, deceleration, maximal impact speed) were transformed a common scale by mapping the values onto the interval $[0.1, 1]$ before calculating the severity sampling scheme.

\subsection{Latin hypercube sampling} 

Latin hypercube sampling was implemented as in \citet{Meng2021}, stratified by cases (i.e., by the original crash events). In brief, the input parameters (off-road glance duration and maximal deceleration during braking) were first transformed to the unit square $[0, 1]^2$ by dividing by the maximum value. For each of the 44 cases, $m$ points were generated at random according to a Latin square design on $[0, 1]^2$. These points were then matched to the closest point in the design space. The sample size was varied from $m=1, \ldots, 46$ observations per case, producing subsamples of $n=44, 88, \ldots, 2024$ observations in total.

\subsection{Leverage sampling} 

Leverage sampling \citep{Ma2015, Ma2022} utilizes importance sampling with probabilities proportional to the statistical leverage scores $\{h_{i}\}_{i=1}^N$ of a linear model with covariates $\{\boldsymbol{z}_i\}_{i=1}^N$. Using weighted least squares, the leverage scores $h_i$ are given by the diagonal elements of the 'hat matrix'
\[
    \mathbf{H} = \mathbf{W}^{1/2}\mathbf{Z}(\mathbf{Z}^{T}\mathbf{W}\mathbf{Z})^{-1}\mathbf{Z}^{T}\mathbf{W}^{1/2},
\]
where $\mathbf{W}$ is a diagonal weight matrix and $\mathbf{Z}$ a design matrix with rows $(1, \mathbf{z}^T)$ \citep{Pregibon1981}. In our application, the elements of $\mathbf{W}$ were chosen equal to the prior scenario weights $p_i$, and $\mathbf{z}_i$ as the off-road glance duration and maximal deceleration during braking for an instance $i$. Note that the statistical leverage scores only depend on the weights and covariates, not the outcomes, and hence could be calculated without additional information about the outcomes of the simulations.

\subsection{Gaussian process active learning} 

Gaussian process active learning was initialized using simple random sampling. New samples were selected at random using probabilistic uncertainty sampling, i.e., with probabilities proportional to the posterior uncertainty of the predictions (standard deviation). This is relatively fast and promotes exploration of the design space, particularly in regions with high prediction uncertainty. Constant predictions were used (i.e., the observed sample mean) if the model could not be fitted. Both subsampling and model fitting were stratified by cases, i.e., performed separately for each original crash event. In each iteration, one sample per case was selected, corresponding to a total batch size of $44$ observations per iteration. Estimation was conducted according to \eqref{eq:theta_application}, replacing the true values by their corresponding predictions.

\subsection{Software}

All sampling methods and experiments were implemented using the \texttt{R} language and environment for statistical computing, version 4.2.1 \citep{R_core_team}. Random forests were implemented using the \texttt{ranger} R-package version 0.14.1 \citep{ranger}, extreme gradient boosting using \texttt{xgboost} version 1.7.5.1, and k-nearest neighbors using \texttt{caret} version 6.0-94 \citep{caret}. \texttt{caret} was also used for hyper-parameter tuning. Latin hypercube sampling was implemented using the \texttt{lhs} package version 1.1.6, with matching using \texttt{MatchIt} version 4.5.5 \citep{MatchIt}. Gaussian process regression was implemented using \texttt{kernlab} version 0.9-32 \citep{kernlab}. The complete \texttt{R} code for the active sampling algorithm, simulation experiments, and data are available online at \href{https://github.com/imbhe/ActiveSampling}{\nolinkurl{https://github.com/imbhe/ActiveSampling}}.

\clearpage
\section{Supplemental Figures}
\label{sec:Supplemental Figures}

\subsection{Additional results: Simulation experiments}
\label{appendix:supplemental_figures_experiments}

\begin{figure}[htb!]
    \centering
    \includegraphics[scale = 1]{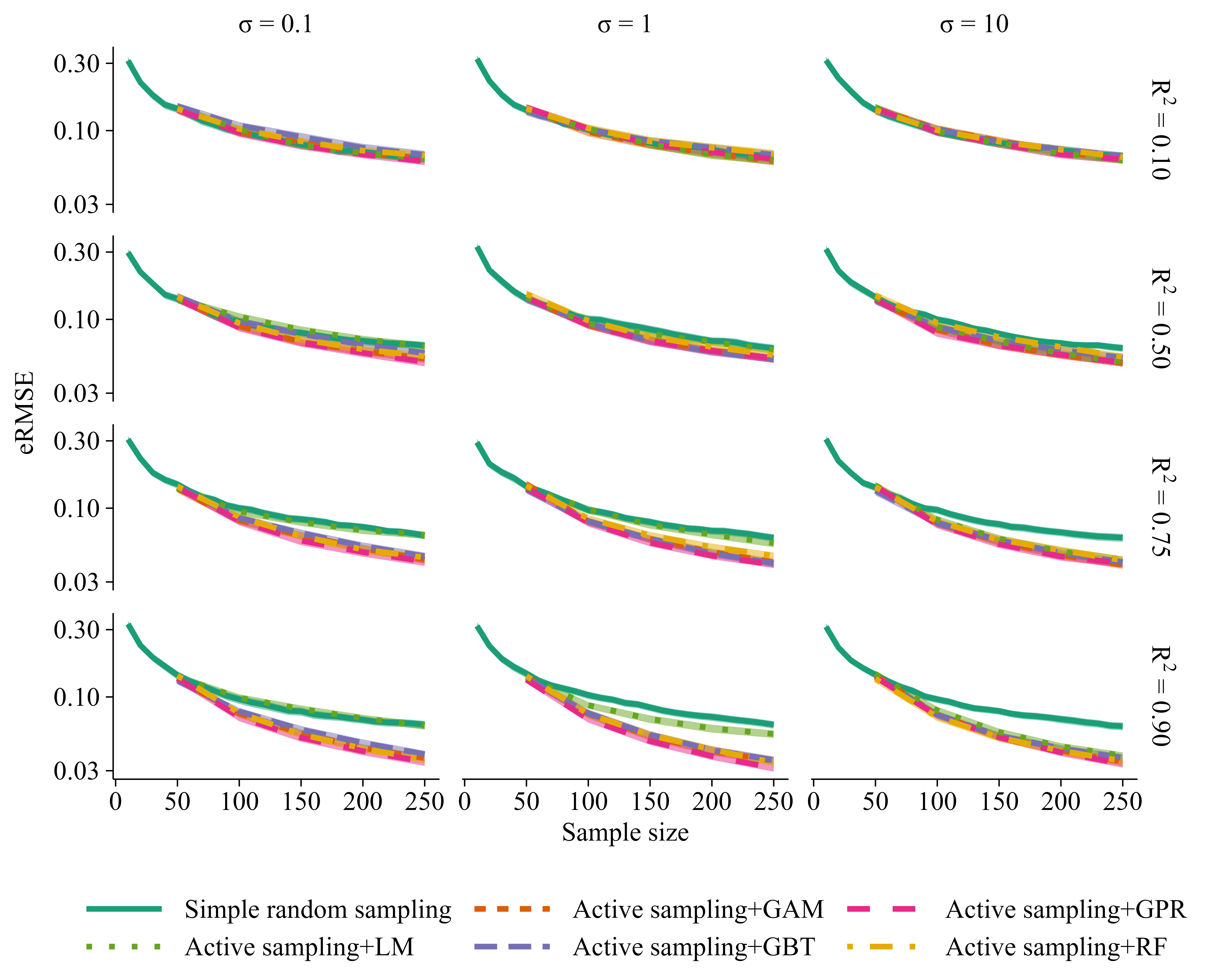}
    \caption{Performance of active sampling using a linear model (LM), generalized additive model (GAM), Gaussian process regression (GPR) or random forests (RF) surrogate model, compared to simple random sampling. The curves show the root mean square errors (eRMSEs) for estimating a finite population mean in a strictly positive scenario (all $y_i>0$) using a linear estimator ($h(u) = u / N$) and batch size $n_k=50$. Shaded regions are 95\% confidence intervals for the eRMSEs based on 500 repeated subsampling experiments.}
    \label{fig:all_ml_models}
\end{figure}

\begin{figure}[htb!]
    \centering
    \includegraphics[scale = 1]{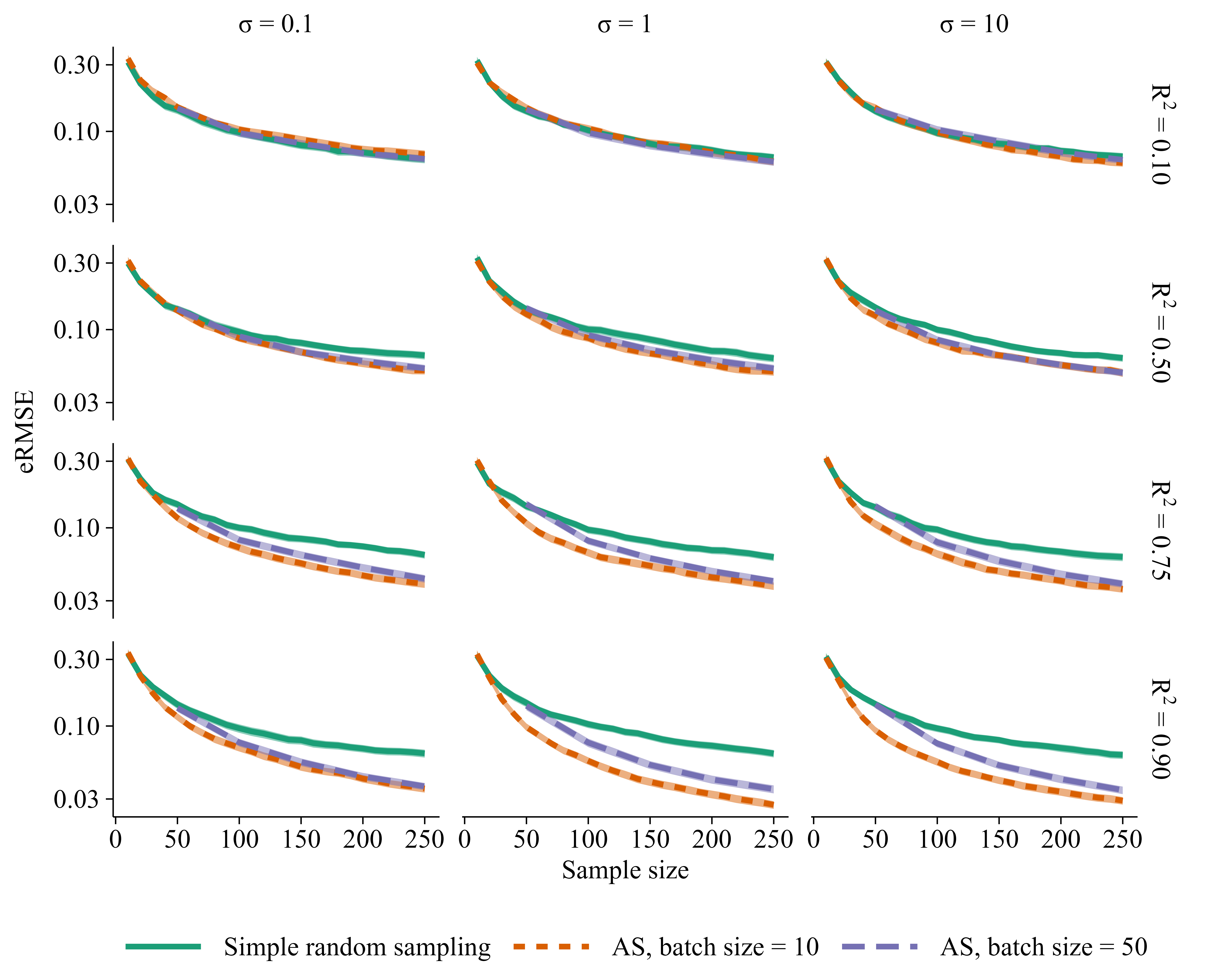}
    \caption{Performance of active sampling (AS) using a generalized additive surrogate model with a batch size of $n_k=10$ or $50$ samples per iteration, compared to simple random sampling. The curves show the root mean square errors (eRMSEs) for estimating a finite population mean in a strictly positive scenario (all $y_i>0$) using a linear estimator ($h(u) = u / N$). Shaded regions are 95\% confidence intervals for the eRMSEs based on 500 repeated subsampling experiments.}    
    \label{fig:batch_size}
\end{figure}

\begin{figure}[htb!]
    \centering
    \includegraphics[scale = 1]{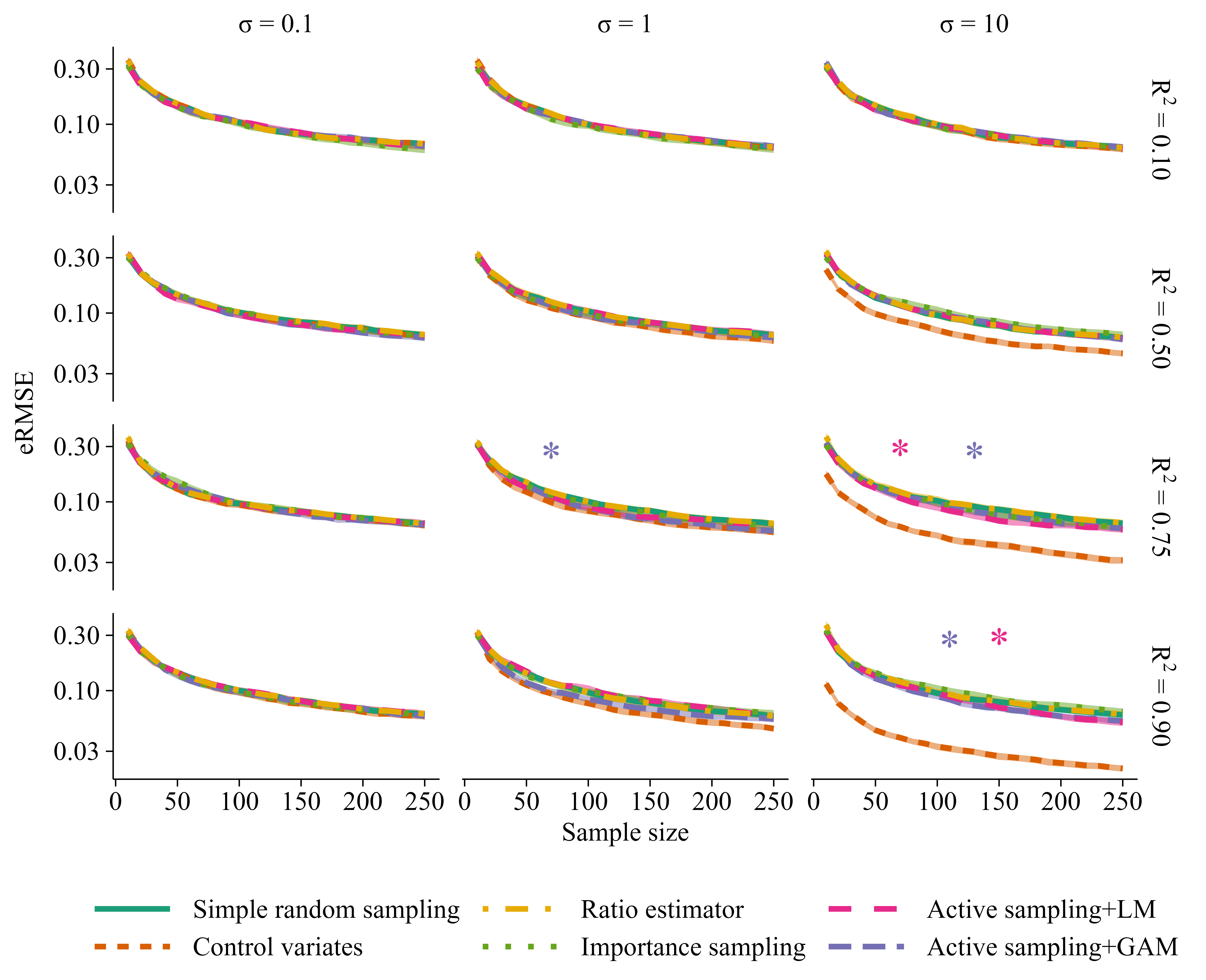}
    \caption{Performance of active sampling using a linear surrogate model (LM) or generalized additive surrogate model (GAM) compared to simple random sampling, ratio estimator, control variates and importance sampling for estimating a finite population mean in a non-restricted scenario ($y_i \in \mathbb{R}$) using a linear estimator ($h(u) = u / N$) and batch size $n_k=10$. The curves and shaded regions are the root mean squared errors (eRMSEs) of the estimators and 95\% confidence intervals for the eRMSEs, respectively, based on 500 repeated subsampling experiments. Asterisks show the smallest sample sizes for which there were persistent significant improvements ($p<.05$) with active sampling compared to simple random sampling.}
    \label{fig:zero_mean}
\end{figure}

\begin{figure}[htb!]
    \centering
    \includegraphics[scale = 1]{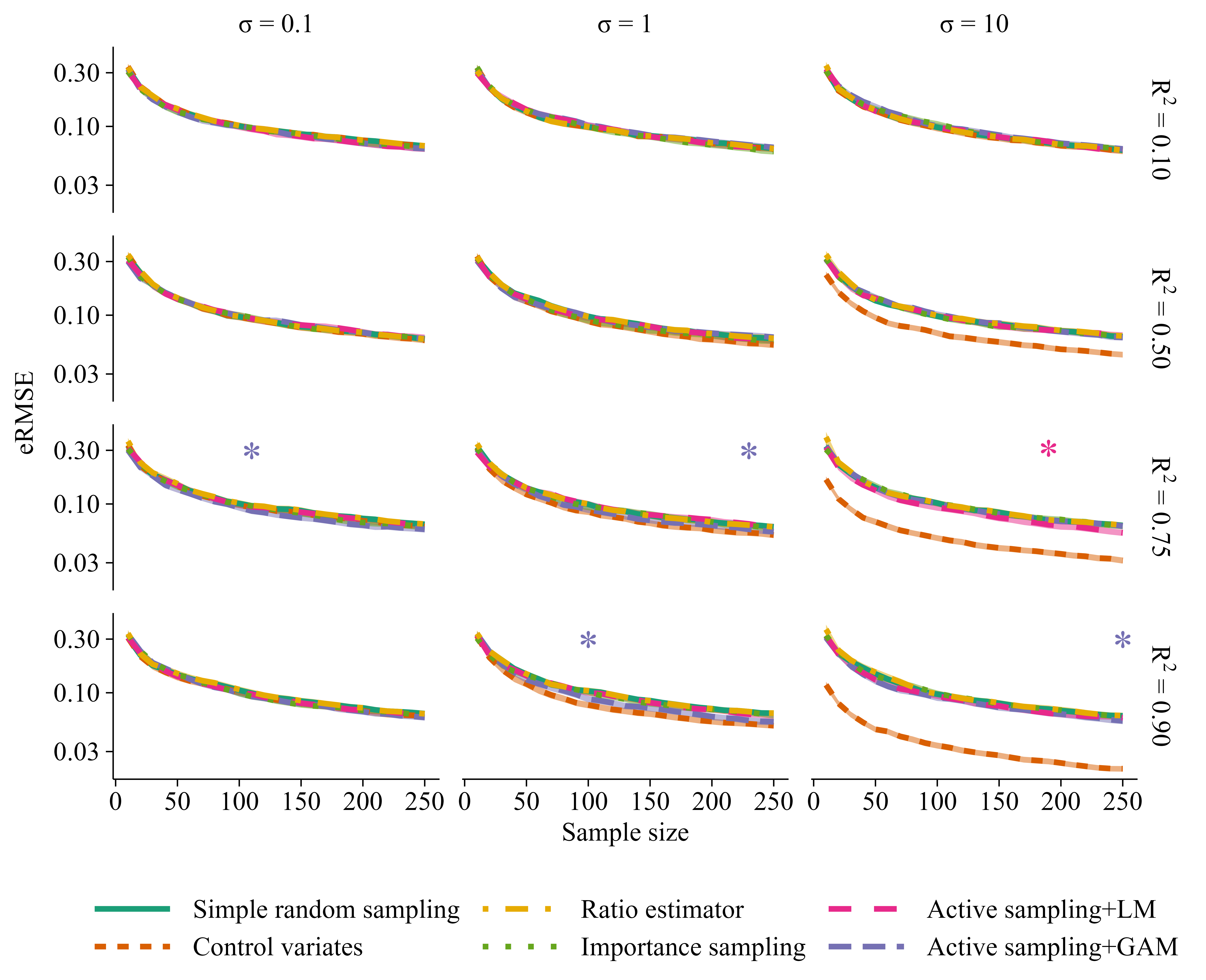}
    \caption{Performance of active sampling using a linear surrogate model (LM) or generalized additive surrogate model (GAM) compared to simple random sampling, ratio estimator, control variates and importance sampling for estimating a finite population mean in a non-restricted scenario ($y_i \in \mathbb{R}$) using a non-linear estimator ($h(\boldsymbol{u}) = u_2 / u_1$, $\boldsymbol{y}_i = (1, y_i)^T$) and batch size $n_k=10$. The curves and shaded regions are the root mean squared errors (eRMSEs) of the estimators and 95\% confidence intervals for the eRMSEs, respectively, based on 500 repeated subsampling experiments. Asterisks show the smallest sample sizes for which there were persistent significant improvements ($p<.05$) with active sampling compared to simple random sampling.}
    \label{fig:hajek}
\end{figure}

\begin{figure}[htb!]
    \centering
    \includegraphics[scale = 1]{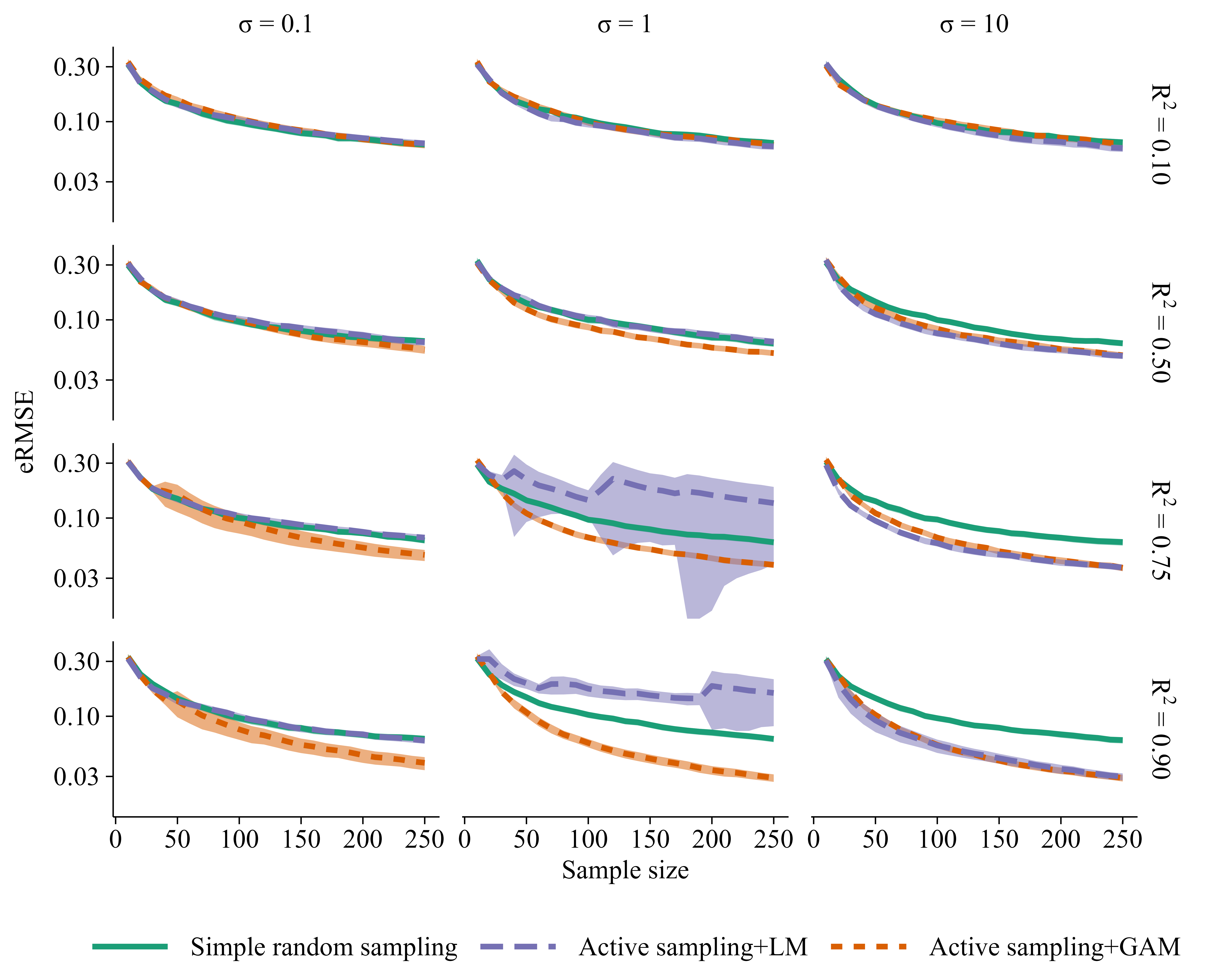}
    \caption{Performance of naive active sampling (ignoring prediction uncertainty) using a linear surrogate model (LM) or generalized additive surrogate model (GAM) compared to simple random sampling for estimating a finite population mean in a strictly positive scenario ($y_i > 0$) using a linear estimator ($h(u) = u / N$) and batch size $n_k=10$. The curves and shaded regions are the root mean squared errors (eRMSEs) of the estimators and 95\% confidence intervals for the eRMSEs, respectively, based on 500 repeated subsampling experiments.}    
    \label{fig:naive_strictly_positive}
\end{figure}

\begin{figure}[htb!]
    \centering
    \includegraphics[scale = 1]{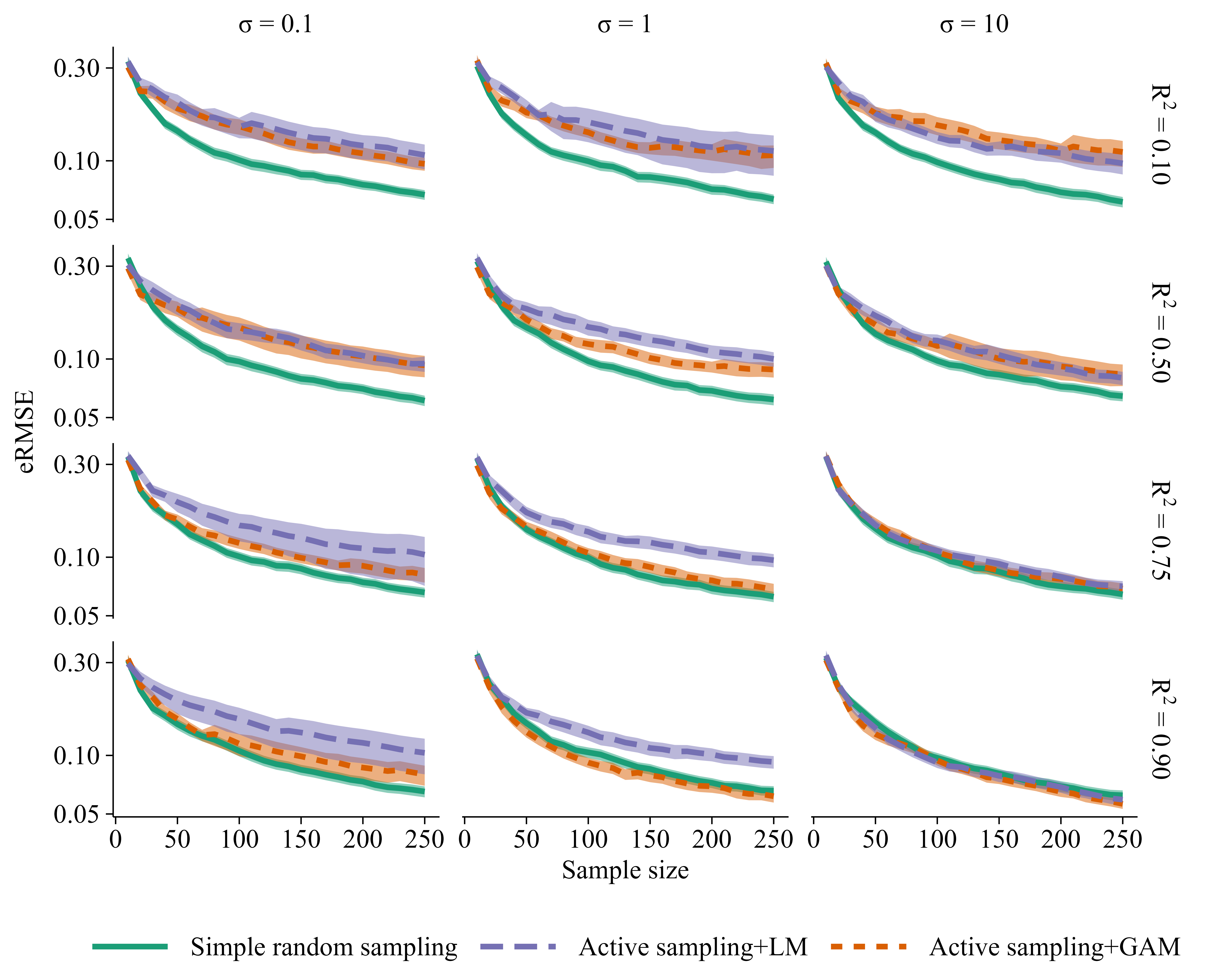}
    \caption{Performance of naive active sampling (ignoring prediction uncertainty) using a linear surrogate model (LM) or generalized additive surrogate model (GAM) compared to simple random sampling for estimating a finite population mean in a non-restricted scenario ($y_i \in \mathbb{R}$) using a non-linear estimator ($h(\boldsymbol{u}) = u_2 / u_1$, $\boldsymbol{y}_i = (1, y_i)^T$) and batch size $n_k=10$. The curves and shaded regions are the root mean squared errors (eRMSEs) of the estimators and 95\% confidence intervals for the eRMSEs, respectively, based on 500 repeated subsampling experiments.}    
    \label{fig:naive_hajek}
\end{figure}

\FloatBarrier
\clearpage

\subsection{Additional results: Application}
\label{appendix:supplemental_figures_application}

\FloatBarrier
\begin{figure}[htb!]
    \begin{center}
        \includegraphics[scale = 1]{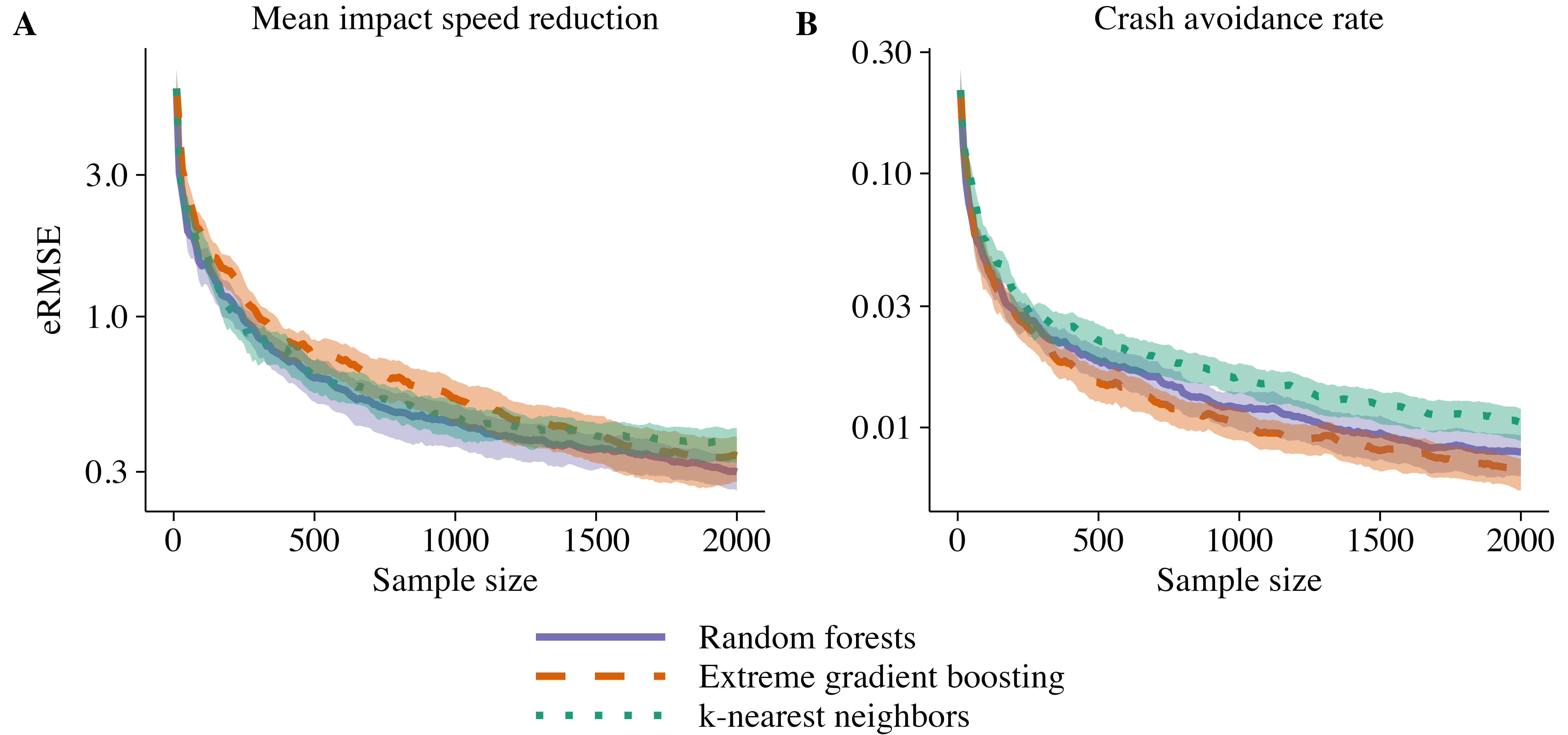}
    \end{center}
    \caption{Root mean squared error (eRMSE) for estimating the mean impact speed reduction (A) and crash avoidance rate (B) using three different machine learning models in the active sampling algorithm: random forests, extreme gradient boosting, and k-nearest neighbors. Shaded regions are 95\% confidence intervals for the eRMSE based on 100 repeated subsampling experiments.}    \label{fig:different_prediction_models_active_sampling_vs_importance_sampling}
\end{figure}

\end{document}